\documentclass[preprint,showpacs,preprintnumbers,amsmath,amssymb]{revtex4}

\usepackage{graphicx}
\usepackage{dcolumn}
\usepackage{bm}

\begin{document}

\title{Kinetic Theory of a Dilute Gas System under Steady Heat Conduction}

\author{Kim Hyeon-Deuk \footnote{kim@yuragi.jinkan.kyoto-u.ac.jp}}

\affiliation{Graduate School of Human and Environmental Studies, Kyoto
University, Kyoto 606-8501, Japan}

\author{Hisao Hayakawa \footnote{hisao@yuragi.jinkan.kyoto-u.ac.jp}}

\affiliation{Department of Physics, Yoshida-South Campus,
  Kyoto University, Kyoto 606-8501, Japan}

\date{\today}

\begin{abstract}
The velocity distribution function of the steady-state Boltzmann equation for
 hard-core molecules in the presence of a temperature gradient has been obtained explicitly to second order in density and the temperature gradient. 
Some thermodynamical quantities are calculated from the velocity
 distribution function for hard-core molecules and compared with those
 for Maxwell molecules and the steady-state Bhatnagar-Gross-Krook(BGK) equation. 
We have found qualitative differences between hard-core molecules
 and Maxwell molecules in the thermodynamical quantities, and also
 confirmed that the steady-state BGK equation belongs to the same
 universality class as Maxwell molecules. 
\end{abstract}


\maketitle

\newpage

\section{Introduction}
The kinetic theory has long history and various investigations have
been carried out analytically, numerically and experimentally.\cite{chapman,mac,resi,kogan,cercignani1,hand}   
Text books on the Boltzmann equation\cite{leb,cohen,cercignani2,cercignani21,arist} 
show the fact that the Boltzmann equation has attracted much 
interest among researchers of kinetic phenomena. 
It is well accepted that the Boltzmann equation is one of the most reliable
kinetic models for describing nonequilibrium phenomena. 

A number of studies on the Boltzmann equation have been based on two
representative models of molecules: hard-core molecules and Maxwell molecules\cite{maxwell0,maxwell01,maxwell1}.   
Maxwell molecules interact via a 
force that is inversely proportional to the fifth power of the distance
$r$, that is, Maxwell molecules interact via a central potential proportional to $r^{-4}$. 
It is well known that calculations involving the Boltzmann equation for
Maxwell molecules become much easier than those for hard-core molecules.\cite{chapman,maxwell0,maxwell01,maxwell1,burnett,maxwell,russia} 
In order to obtain the velocity distribution function of a dilute gas
system in a nonequilibrium state, 
various methods, such as the Chapman-Enskog method, the Grad
method and the Hilbert method, have been presented.\cite{chapman,mac,resi,kogan,cercignani1,hand,leb,cohen,cercignani2,cercignani21,arist,maxwell1,burnett,maxwell,russia,grad,grad1,grad11}  
They are well known as methods to derive the normal solutions of the Boltzmann
equation. 

Other kinetic models besides the Boltzmann equation, such as the linearized Boltzmann
equation\cite{cercignani1,cercignani} and the Bhatnagar-Gross-Krook
(BGK) equation\cite{bgk,lebo,korea,korea1,santos3,santos31,santos4,santos41,santos42,santos2,santos21,santos22}, have also been
proposed in order to avoid mathematical difficulties in dealing with the
collision term of the Boltzmann equation.   
It has been believed that results of those kinetic models approximately
agree with those of the Boltzmann equation, though they are accepted as 
quantitatively different ones from the Boltzmann equation:   
\textit{e.g.}
 the Prandtl number is $2/3$ for the Boltzmann equation, while it is $1$ for
the BGK equation.\cite{korea,korea1}
The BGK equation retains important properties of the Boltzmann
equation, such as the conservation laws and
the H-theorem. 
It has been felt that characteristics of molecules, such as hard-core molecules and 
Maxwell molecules, can be absorbed into the relaxation time $\tau$.   
The most important contribution concerning the solution of the BGK
equation has been made by Santos and his coworkers.\cite{santos3,santos31,santos4,santos41,santos42,santos2,santos21,santos22}  
They have solved the steady-state BGK equation exactly and compared its
exact solution with
the Chapman-Enskog solution derived to arbitrary order.\cite{santos3,santos31,santos4,santos41,santos42} 
They have concluded that the Chapman-Enskog solution is asymptotically correct, although the Chapman-Enskog series diverge.\cite{santos3,santos31,santos4,santos41,santos42} 
They have also compared high order velocity distribution
function for the steady-state BGK equation with that for the Boltzmann equation for
Maxwell molecules under \textit{uniform} shear flow.\cite{santos2,santos21,santos22} 

On the other hand, it had been believed that Burnett\cite{burnett} determined the complete second order solution of the Boltzmann
equation by the Chapman-Enskog method.\cite{chapman}
Importance of the second-order coefficients has been demonstrated for
descriptions of shock wave profiles and sound
propagation phenomena.\cite{foch,shocksound,shocksound1,shocksound2} 
However, we have realized that Burnett's solution\cite{burnett} is not
 complete. 
Though there are various studies on the transport coefficients of the Boltzmann
equation to second order\cite{chapman,resi,kogan,pop,grad1,grad11,wc} or even to third
order\cite{foch}, 
we have found that nobody has derived the explicit velocity distribution
function of the Boltzmann
equation for hard-core molecules to second order. 
This is a result of mathematical difficulties, as was indicated by Fort and Cukrowski\cite{fort}.     
For Maxwell molecules, Schamberg\cite{maxwell} has derived the
precise velocity distribution function of the Boltzmann
equation to second order by the Chapman-Enskog method, while Shavaliev\cite{russia} has derived
it implicitly by the moment method. 

In this paper, we derive the
explicit velocity distribution function of the steady-state Boltzmann
equation for hard-core molecules to second order in density and temperature gradients by the Chapman-Enskog method. 
The derivation of it is of some physical significance. 
For example, it has been required for calculating a nonequilibrium effect on
the rate of chemical reaction\cite{fort}, because the nonequilibrium
effect on the rate of chemical reaction
does not appear to first order and it cannot be calculated by the moment 
method.\cite{kim1}
There also exists a need to confirm the existence of universal nonlinear
nonequilibrium statistical mechanics 
from the microscopic
viewpoint as mentioned in ref.\cite{fort}, so that it is necessary to derive the
explicit velocity distribution function of the steady-state Boltzmann
equation for hard-core molecules to second order and compare it with that for Maxwell molecules. 
Additionally, the derivation of the explicit second-order solution of the steady-state Boltzmann
equation for hard-core molecules by the Chapman-Enskog method
will contribute to an understanding of the difference or the
correspondence 
between the Chapman-Enskog method and the Grad method by the direct
comparison of the solution by the Chapman-Enskog method with that by the
Grad method. 
In the present paper, we also discuss the relation between the steady-state Boltzmann
equation and the steady-state BGK equation which has not been fully
understood yet. 

The organization of this paper is as follows. 
In \S \ref{solve}, we will introduce the Chapman-Enskog method to solve the steady
state Boltzmann equation. 
In \S \ref{f2}, we will derive the explicit form of the velocity distribution
function of the steady-state Boltzmann
equation for hard-core molecules to second order in the density and
temperature gradients. 
In particular, the result for the first-order coefficients is shown in
eq.(\ref{be37}), and the results for the second-order
coefficients are shown in eqs.(\ref{be42}) and (\ref{be45}). 
The velocity
distribution function of the steady-state Boltzmann
equation to second order for hard-core molecules is shown explicitly in
eq.(\ref{be46}) and graphically in Fig.\ref{f2cxcy}, and compared directly with that for Maxwell
molecules (\ref{be465}) in Fig.\ref{f2hikaku}. 
In \S \ref{application}, we will apply the velocity distribution
function to second order for hard-core molecules (\ref{be46}) and those for Maxwell
molecules (\ref{be465}) and the steady-state BGK equation to a nonequilibrium steady-state system under steady heat conduction and calculate
some thermodynamical quantities. 
We stress the existence of the qualitative differences among hard-core
molecules, Maxwell molecules and the steady-state BGK equation in the pressure tensor (\ref{be51})
 and the kinetic temperature (\ref{be54}). 
Our discussion and conclusion are written in \S \ref{discussion}
and \ref{conclusion}, respectively. 

\section{Method for Solving the Steady-State Boltzmann Equation}\label{solve}
\begin{figure}[htbp]
\begin{center}
\includegraphics[width=10.cm]{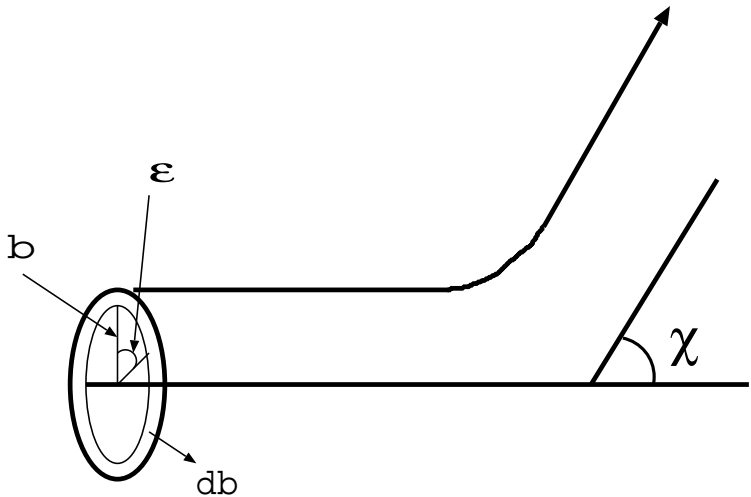} 
\caption{Schematic description of an interaction.}
\label{interaction}
\end{center}
\end{figure}

Let us introduce the Chapman-Enskog method to solve the steady-state Boltzmann equation in this section. 
Assume that we have a system of dilute gases in a steady state, with velocity distribution
function $f_{\mathrm{1}}=f({\bf r},{\bf v}_{\mathrm{1}})$. 
The appropriate steady-state Boltzmann equation is 
\begin{equation}  
{\bf v}_{\mathrm{1}}\cdot\nabla f_{\mathrm{1}}=
J(f_{\mathrm{1}},f_{\mathrm{2}}),\label{be1}
\end{equation}
where the collision integral $J(f_{\mathrm{1}},f_{\mathrm{2}})$ is expressed as 
\begin{equation}  
J(f_{\mathrm{1}},f_{\mathrm{2}})=
\int \int \int (f^{\prime}_{\mathrm{1}}f^{\prime}_{\mathrm{2}}-f_{\mathrm{1}}f_{\mathrm{2}})g b \mathrm{d}b \mathrm{d}\epsilon \mathrm{d}{\bf v}_{\mathrm{2}},\label{be2}
\end{equation}
with $f_{\mathrm{1}}^{\prime}=f({\bf r},{\bf v}_{\mathrm{1}}^{\prime})$
and $f_{\mathrm{2}}^{\prime}=f({\bf r},{\bf v}_{\mathrm{2}}^{\prime})$:  
${\bf v}_{\mathrm{1}}^{\prime}$ and ${\bf v}_{\mathrm{2}}^{\prime}$ are
postcollisional velocities of ${\bf v}_{\mathrm{1}}$ and ${\bf v}_{\mathrm{2}}$, respectively. 
The relative velocity of two molecules before and after the
interaction has the same magnitude $g=|{\bf v}_{\mathrm{1}}-{\bf
v}_{\mathrm{2}}|$
; the angle between the directions of the relative
velocity before and after the interaction is represented by $\chi$. 
The relative position of the two molecules is represented by $b$, called
the impact parameter, and $\epsilon$ represents the orientation of
the plane in which ${\bf g}$ and ${\bf g^{\prime}}={\bf v}_{\mathrm{1}}^{\prime}-{\bf
v}_{\mathrm{2}}^{\prime}$ belong.    
The impact parameter $b$ depends on the kind of interactions between molecules. 
Note that $\chi$ can be expressed as a function of $b$ for a central force. 
(see Fig.\ref{interaction})

Suppose that the velocity distribution function $f_{\mathrm{1}}$ can be expanded as:
\begin{equation}  
f_{\mathrm{1}}=f^{(0)}_{\mathrm{1}}+f^{(1)}_{\mathrm{1}}+f^{(2)}_{\mathrm{1}}+\cdots=f^{(0)}_{\mathrm{1}}(1+\phi^{(1)}_{\mathrm{1}}+\phi^{(2)}_{\mathrm{1}}+\cdots). \label{be3}
\end{equation} 
$f^{(0)}_{\mathrm{1}}$ is the local Maxwellian distribution function, written as 
\begin{equation}  
f^{(0)}_{\mathrm{1}}=n({\bf r})\left(\frac{m}{2\pi \kappa T({\bf r})}\right)^{\frac{3}{2}}\exp\left[-\frac{m{\bf v}_{\mathrm{1}}^{2}}{2\kappa T({\bf r})}\right],\label{be3.5} 
\end{equation} 
with $m$ mass of the molecules and $\kappa$ the Boltzmann constant. 
$n({\bf r})$ and $T({\bf r})$ will be identified later as the density and the temperature at position ${\bf r}$, respectively. 
Substituting eq.(\ref{be3}) into the steady-state Boltzmann equation (\ref{be1}), we
arrive at the following set of equations which we will solve completely
in this paper:  
\begin{eqnarray}  
L[f^{(0)}_{\mathrm{1}}]\phi^{(1)}_{\mathrm{1}}={\bf v}_{\mathrm{1}}\cdot\nabla f^{(0)}_{\mathrm{1}}, \label{be4}
\end{eqnarray}
to first order and
\begin{eqnarray}  
L[f^{(0)}_{\mathrm{1}}]\phi^{(2)}_{\mathrm{1}}={\bf v}_{\mathrm{1}}\cdot\nabla f^{(1)}_{\mathrm{1}}-J(f^{(1)}_{\mathrm{1}},f^{(1)}_{\mathrm{2}}),\label{be5}
\end{eqnarray}
to second order. 
The linear integral operator $L[f^{(0)}_{\mathrm{1}}]$ is defined as
\begin{eqnarray}  
L[f^{(0)}_{\mathrm{1}}]X_{\mathrm{1}}\equiv\int \int \int f^{(0)}_{\mathrm{1}}f_{\mathrm{2}}^{(0)}(X_{\mathrm{1}}^{\prime}-X_{\mathrm{1}}+X^{\prime}_{\mathrm{2}}-X_{\mathrm{2}})g b \mathrm{d}b \mathrm{d}\epsilon \mathrm{d}{\bf v}_{\mathrm{2}}. \label{be6}
\end{eqnarray}
The solubility conditions of the integral equation (\ref{be4}) are given by
\begin{eqnarray}  
\int \Phi_{\mathrm{i}} {\bf v}_{\mathrm{1}}\cdot\nabla f^{(0)}_{\mathrm{1}} \mathrm{d}{\bf v}_{\mathrm{1}}=0,\label{be7}
\end{eqnarray}
where $\Phi_{\mathrm{i}}$ is one of the collisional invariants:
\begin{eqnarray}  
\Phi_{1}=1, \quad \Phi_{2}=m{\bf v}_{\mathrm{1}}, \quad \Phi_{3}=\frac{1}{2}m{\bf v}_{\mathrm{1}}^{2}.\label{be8}
\end{eqnarray}
Substituting eq.(\ref{be3.5}) into the solubility conditions (\ref{be7}),
it is seen that $n\kappa T$ is uniform in the steady state. 
We use this result in our calculation. 
Similarly, the solubility conditions of the integral equation (\ref{be5}) are given by
\begin{eqnarray}  
\int \Phi_{\mathrm{i}} {\bf v}_{\mathrm{1}}\cdot\nabla f^{(1)}_{\mathrm{1}} \mathrm{d}{\bf v}_{\mathrm{1}}=0.
\label{be9}
\end{eqnarray}

To construct solutions of the integral equations
(\ref{be4}) and (\ref{be5}) definite, five further conditions must be
specified; we identify the density:   
\begin{equation}  
n({\bf r})\equiv\int f_{\mathrm{1}} \mathrm{d}{\bf v}_{\mathrm{1}}=\int f^{(0)}_{\mathrm{1}} \mathrm{d}{\bf v}_{\mathrm{1}},\label{be10}
\end{equation} 
the temperature:  
\begin{equation}  
\frac{3n({\bf r})\kappa T({\bf r})}{2}\equiv
\int \frac{m{\bf v}_{\mathrm{1}}^{2}}{2}f_{\mathrm{1}} \mathrm{d}{\bf v}_{\mathrm{1}}=\int \frac{m{\bf v}_{\mathrm{1}}^{2}}{2}f^{(0)}_{\mathrm{1}} \mathrm{d}{\bf v}_{\mathrm{1}}, \label{be11}
\end{equation}
and the mean flow: 
\begin{equation}  
{\bf C}_{0}\equiv\int m {\bf v}_{\mathrm{1}}f_{\mathrm{1}} \mathrm{d}{\bf v}_{\mathrm{1}}=\int m {\bf v}_{\mathrm{1}}f^{(0)}_{\mathrm{1}} \mathrm{d}{\bf v}_{\mathrm{1}}.\label{be12}
\end{equation} 
Here we assume that no mean flow, i.e. ${\bf C}_{0}=0$, exists in
the system. 
The introduction of these conditions distinguishes the Chapman-Enskog
adopted here from the Hilbert method in which the conserved quantities
are also expanded.\cite{resi} 
We assert that conditions (\ref{be10}), (\ref{be11}) and (\ref{be12}) do not
affect all our results in this paper. 
It should be noted that, to solve the integral equations (\ref{be4})
and (\ref{be5}), we should
consider only the case in which the right-hand sides of eqs.(\ref{be4})
and (\ref{be5}) are not zero: if the right-hand sides of eqs.(\ref{be4})
and (\ref{be5}) are zero, the integral equations (\ref{be4})
and (\ref{be5}) become homogeneous equations which do not have any
particular solutions.\cite{resi} 

\section{Burnett's Method}\label{f2}
\subsection{A general form of the velocity distribution function}
To solve the integral equations (\ref{be4}) and (\ref{be5}),
Burnett\cite{burnett} assumed a general form of the velocity
distribution function: 
\begin{eqnarray}  
f_{\mathrm{1}}=f^{(0)}_{\mathrm{1}}[\sum_{r=0}^{\infty} r! \Gamma(r+\frac{3}{2}) B_{0r}S^{r}_{\frac{1}{2}}({\bf c}_{\mathrm{1}}^{2})+\sum_{k=1}^{\infty}\left( \frac{m}{2\kappa T} \right)^{\frac{k}{2}}\sum_{r=0}^{\infty} r! \Gamma(k+r+\frac{3}{2})Y_{kr}({\bf c}_{\mathrm{1}})S^{r}_{k+\frac{1}{2}}({\bf c}_{\mathrm{1}}^{2})].\label{be13}
\end{eqnarray}
Here ${\bf c}_{\mathrm{1}}\equiv (m/2\kappa T)^{1/2}{\bf v}_{\mathrm{1}}$
is the scaled velocity and $\Gamma(X)$ represents the Gamma function. 
$S_{k}^{p}(X)$ is a Sonine
polynomial, defined by 
\begin{eqnarray}  
(1-\omega)^{-k-1}e^{-\frac{X\omega}{1-\omega}}=\sum_{p=0}^{\infty} \Gamma(p+k+1) S_{k}^{p}(X)\omega^p.  
\label{be13.5}
\end{eqnarray}
$Y_{kr}({\bf c}_{\mathrm{1}})$ is a linear combination of spherical 
harmonic functions: 
\begin{eqnarray}  
Y_{kr}({\bf c}_{\mathrm{1}})=B_{kr}Y_{k}({\bf c}_{\mathrm{1}})+2\sum_{p=1}^{k}\frac{(k-p)!}{(k+p)!}[B_{kr}^{(p)}Y_{k}^{(p)}({\bf c}_{\mathrm{1}})+C^{(p)}_{kr}Z_{k}^{(p)}({\bf c}_{\mathrm{1}})], \label{be14}
\end{eqnarray}
where $B_{kr}$, $B_{kr}^{(p)}$ and $C^{(p)}_{kr}$ are coefficients to be
determined. 
Introducing the normal spherical coordinate representation for ${\bf
c}_{{\mathrm{1}}}$, i.e. $c_{{\mathrm{1}}x}=c_{{\mathrm{1}}}\sin \theta
\cos\phi$, $c_{{\mathrm{1}}y}=c_{{\mathrm{1}}}\sin \theta
\sin\phi$ and $c_{{\mathrm{1}}z}=c_{{\mathrm{1}}}\cos \theta$,  
the spherical harmonic functions $Y_{k}({\bf c}_{\mathrm{1}}),Y_{k}^{(p)}({\bf c}_{\mathrm{1}})$
and $Z_{k}^{(p)}({\bf c}_{\mathrm{1}})$ are expressed as
\begin{eqnarray}
Y_{k}({\bf c}_{\mathrm{1}})= \left( \frac{2\kappa T}{m} \right)^{\frac{k}{2}}c_{\mathrm{1}}^{k}P_{k}(\cos\theta),\label{be15}
\end{eqnarray}
and
\begin{eqnarray}
Y_{k}^{(p)}({\bf c}_{\mathrm{1}})=(-1)^{p}\left( \frac{2\kappa T}{m} \right)^{\frac{k}{2}} c_{\mathrm{1}}^{k}P^{(p)}_{k}(\cos\theta) \cos p\phi, \label{be16} 
\end{eqnarray}
and
\begin{eqnarray}
Z_{k}^{(p)}({\bf c}_{\mathrm{1}})=(-1)^{p}\left(\frac{2\kappa T}{m} \right)^{\frac{k}{2}}c_{\mathrm{1}}^{k}P^{(p)}_{k}(\cos\theta) \sin p\phi, \label{be17}
\end{eqnarray}
with the Legendre polynomial $P_{k}(\cos\theta)$ and the associated Legendre
polynomial $P_{k}^{(p)}(\cos\theta)$. 

Assumption of the velocity distribution function of eq.(\ref{be13}) has some mathematical advantages in our
calculation.  
Firstly, it is sufficient to determine the coefficients
$B_{kr}$, because $B_{kr}^{(p)}$ and $C^{(p)}_{kr}$ can be always determined from
$B_{kr}$ by a transformation of axes owing to the properties of the
spherical harmonic functions (\ref{be15}), (\ref{be16}) and (\ref{be17}). 
We call $B_{kr}$, $B_{kr}^{(p)}$ and $C^{(p)}_{kr}$ the family of
$B_{kr}$. 
Secondly, some important physical quantities are related to 
coefficients $B_{kr},B_{kr}^{(p)},C^{(p)}_{kr}$: 
\textit{e.g.} the density (\ref{be10}), the temperature (\ref{be11}) and the mean flow
(\ref{be12}) with $f_{\mathrm{1}}$ in
eq.(\ref{be13}) lead to the five equivalent
conditions\cite{burnett,maxwell}: 
\begin{eqnarray}  
B_{00}=1,\quad B_{10}=B^{(1)}_{10}=C^{(1)}_{10}=0, \quad B_{01}=0. \label{be18}
\end{eqnarray}
Similarly, the pressure tensor $P_{ij}$ defined by
\begin{eqnarray}  
P_{ij}=\left(\frac{2\kappa T}{m}\right)^{\frac{5}{2}}\int^{\infty}_{-\infty}\mathrm{d}{\bf c}_{\mathrm{1}}mc_{{\mathrm{1}} i}c_{{\mathrm{1}} j}f_{\mathrm{1}},\label{be17.5}
\end{eqnarray}
for $i, j=x, y$ and $z$ is related to the family of
$B_{20}$, which is the only family in which Burnett was interested.\cite{burnett}  

The coefficients $B_{kr}$ except for those
in eq.(\ref{be18}) can be calculated as follows. 
Multiplying the steady-state Boltzmann equation (\ref{be1}) by 
\begin{eqnarray}  
Q_{kr}({\bf c}_{\mathrm{1}})\equiv(k+\frac{1}{2})\sqrt{\pi}(\frac{m}{2\kappa T})^{\frac{k}{2}}Y_{k}({\bf c}_{\mathrm{1}})S^{r}_{k+\frac{1}{2}}({\bf c}_{\mathrm{1}}^{2}),\label{be20}
\end{eqnarray}
and then integrating over $(2\kappa T/m)^{1/2}{\bf c}_{\mathrm{1}}$, it is found that 
\begin{eqnarray}  
-\left(\frac{2\kappa T}{m}\right)^{\frac{1}{2}}<{\bf c}_{\mathrm{1}}\cdot\nabla Q_{kr}>_{av}+\nabla\cdot \left[\left(\frac{2\kappa T}{m}\right)^{\frac{1}{2}}<{\bf c}_{\mathrm{1}} Q_{kr}>_{av}\right]=\nonumber \\
\left(\frac{2\kappa T}{m}\right)^{3}\int \int \int \int (Q^{\prime}_{kr}-Q_{kr}) f_{\mathrm{1}}f_{\mathrm{2}}g b \mathrm{d}b \mathrm{d}\epsilon \mathrm{d}{\bf c}_{\mathrm{2}}\mathrm{d}{\bf c}_{\mathrm{1}},
\label{be19}
\end{eqnarray}
where $<X>_{av}$ represents $(2\kappa T/m)^{3/2}\int X f_{\mathrm{1}} \mathrm{d}{\bf
c}_{\mathrm{1}}$, and $Q^{\prime}_{kr}$ represents the postcollisional
$Q_{kr}$. 
We should calculate both sides of eq.(\ref{be19}) for every $k$ and
$r$, because eq.(\ref{be19}) leads to equations to determine $B_{kr}$,
as will be shown in Appendices \ref{a10} and \ref{a50}. 

For convenience, we introduce $\Omega_{kr}$ and $\Delta_{kr}$ as the left-hand 
and right-hand sides of
eq.(\ref{be19}), respectively, i.e.   
\begin{equation}  
\Omega_{kr}\equiv-\left(\frac{2\kappa T}{m}\right)^{\frac{1}{2}}<{\bf c}_{\mathrm{1}}\cdot\nabla Q_{kr}>_{av}+\nabla\cdot \left[\left(\frac{2\kappa T}{m}\right)^{\frac{1}{2}}<{\bf c}_{\mathrm{1}} Q_{kr}>_{av}\right],\label{be19.0}
\end{equation}
and
\begin{equation}  
\Delta_{kr}\equiv\left(\frac{2\kappa T}{m}\right)^{3}\int \int \int \int (Q^{\prime}_{kr}-Q_{kr}) f_{\mathrm{1}}f_{\mathrm{2}}g b \mathrm{d}b \mathrm{d}\epsilon \mathrm{d}{\bf c}_{\mathrm{2}}\mathrm{d}{\bf c}_{\mathrm{1}}.\label{be19.1}
\end{equation}
We will calculate $\Omega_{kr}$ and $\Delta_{kr}$ separately. 
The result of $\Omega_{kr}$ becomes 
\begin{eqnarray}  
\Omega_{kr}&=&\frac{n}{T}(\frac{2\kappa T}{m})^{\frac{1}{2}}[(r+\frac{k}{2})(D_{k,r}\partial_{x} T+E_{k,r}\partial_{y} T+G_{k,r}\partial_{z} T) \nonumber \\
&-&(D_{k,r-1}\partial_{x} T+E_{k,r-1}\partial_{y} T+G_{k,r-1}\partial_{z} T)]\nonumber\\
&+&\partial_{x}\left[n(\frac{2\kappa T}{m})^{\frac{1}{2}}D_{k,r}\right]+\partial_{y} \left[n(\frac{2\kappa T}{m})^{\frac{1}{2}}E_{k,r}\right]+\partial_{z} \left[n(\frac{2\kappa T}{m})^{\frac{1}{2}}G_{k,r}\right], \label{be25.5}
\end{eqnarray}
where $\partial_{\mathrm{i}} X$ denotes $\partial X/\partial \mathrm{i}$ for $\mathrm{i}=x, y$ and $z$.  
$D_{k,r}$, $E_{k,r}$ and $G_{k,r}$ are functions of the family of $B_{kr}$, as is written in Appendix \ref{a3}. 

\subsection{The collision term $\Delta_{\lowercase{kr}}$}\label{delta}
Next, we calculate the collision term $\Delta_{kr}$ in eq.(\ref{be19.1}). 
We should specify the kind of interactions of molecules so as to perform the calculation
of the collision
term $\Delta_{kr}$; the impact parameter $b$ is explicitly determined by
specifying the type of interaction. 
For hard-core molecules, the impact parameter $b$ is given by the relation
\begin{eqnarray}  
b=\sigma \cos \frac{\chi}{2},\label{be26}
\end{eqnarray}
where $\sigma$ is the hard-core molecular diameter and $\chi$ is the 
scattering angle (see Fig.\ref{interaction}).  
Therefore, $\Delta_{kr}$ for hard-core molecules, i.e. $\Delta_{kr}^{\mathrm{H}}$, becomes\cite{burnett}
\begin{eqnarray}
\Delta_{kr}^{\mathrm{H}}&=&\frac{\sigma^{2}}{2}\int_{0}^{\pi} [F_{kr}^{1}(\chi)-F_{kr}^{1}(0) ]\sin\frac{\chi}{2} \cos\frac{\chi}{2} \mathrm{d}\chi, \label{be28}
\end{eqnarray}
where $F_{kr}^{\mu}(\chi)$ is defined as
\begin{eqnarray}
F_{kr}^{\mu}(\chi)\equiv\left(\frac{2\kappa T}{m}\right)^{3}\int \int \int Q^{\prime}_{kr} f_{\mathrm{1}}f_{\mathrm{2}}g^{\mu} \mathrm{d}\epsilon \mathrm{d}{\bf c}_{\mathrm{2}}\mathrm{d}{\bf c}_{\mathrm{1}}, \label{be27}
\end{eqnarray}
and we have used $F_{kr}^{\mu}(0)=(2\kappa T/m)^{3}\int \int\int Q_{kr} f_{\mathrm{1}}f_{\mathrm{2}}g^{\mu} \mathrm{d}\epsilon \mathrm{d}{\bf c}_{\mathrm{2}}\mathrm{d}{\bf c}_{\mathrm{1}}$. 
Note that $\chi=0$ if $b > \sigma$.  
For Maxwell molecules, the impact parameter $b$ is given by the relation
\begin{eqnarray}  
b \mathrm{d}b = \frac{1}{g} H(\chi) \mathrm{d}\chi,\label{be26.00}
\end{eqnarray}
where $H(\chi)$ is a function of $\chi$.\cite{hand,maxwell0,maxwell01,maxwell1} 
Therefore, $\Delta_{kr}$ for Maxwell molecules, i.e. $\Delta_{kr}^{\mathrm{M}}$, becomes 
\begin{eqnarray}
\Delta_{kr}^{\mathrm{M}}&=&\int_{0}^{\pi} [F_{kr}^{0}(\chi)-F_{kr}^{0}(0) ]H(\chi) \mathrm{d}\chi.  \label{be28.00}
\end{eqnarray}
Since $\Delta_{kr}^{\mathrm{M}}$ has been calculated by
Schamberg\cite{maxwell}, we will calculate only
$\Delta_{kr}^{\mathrm{H}}$ in this paper. 
Schamberg's result for $\Delta_{kr}^{\mathrm{M}}$ is briefly summarized
in our web page.\cite{address}  
 
From eq.(\ref{be28}), it is sufficient to calculate $F_{kr}^{1}(\chi)$ for $\Delta_{kr}^{\mathrm{H}}$. 
The details of $F_{kr}^{1}(\chi)$ are written in Appendix \ref{a2}. 
Several explicit forms of $\Delta_{kr}^{\mathrm{H}}$ are also
demonstrated in Appendices \ref{a10} and \ref{a50}. 
From the definitions (\ref{be19.0}) and (\ref{be19.1}), both sides of eq.(\ref{be19}) for arbitrary $k$ and
$r$ can be calculated for hard-core molecules via 
\begin{eqnarray}
\Omega_{kr}^{\mathrm{H}}=\Delta_{kr}^{\mathrm{H}},\label{be0}
\end{eqnarray}
which produces a set of simultaneous equations determining the coefficients
$B_{kr}$, as is explained in Appendices \ref{a10} and \ref{a50}. 
Here $\Omega_{kr}^{\mathrm{H}}$ denotes $\Omega_{kr}$ in
eq.(\ref{be25.5}) for hard-core
molecules. 

\subsection{Determination of $B_{kr}$}\label{coefficient}
We will determine the first-order coefficients
$B_{kr}^{\mathrm{I}}$ and the second-order coefficients
$B_{kr}^{\mathrm{II}}$ in accordance with the previous two subsections, which corresponds to solving the integral equations (\ref{be4})
and (\ref{be5}), respectively. 
Here the upper suffices $\mathrm{I}$ and $\mathrm{II}$ are introduced to
specify the order of $B_{kr}$. 
We have provided an example of our Mathematica program for calculating these
coefficients.\cite{address}  

\subsubsection{The first order}\label{first}
We show the results of the first-order coefficients
 $B_{kr}^{\mathrm{I}}$ of which the
solution of the integral equation (\ref{be4}), $\phi^{(1)}_{\mathrm{1}}$, is composed. 
They can be written in the form: 
\begin{eqnarray}
B_{kr}^{\mathrm{I}}=\delta_{k,1} b_{1r}\frac{15}{16}\frac{\partial_{z} T}{\sqrt{2\pi}\sigma^{2}nT}. \label{be37}
\end{eqnarray}
Values of the constants $b_{1r}$ are given in Table \ref{b1r}. 
The calculation of $B_{kr}^{\mathrm{I}}$ is explained in Appendix
\ref{a10}. 
It is seen that $B_{kr}^{\mathrm{I}}$ is of the order of the Knudsen
number $K$, which means that the mean free path of molecules should be much less than
the characteristic length for changes in the macroscopic
variables. 
Though Burnett derived $B_{kr}^{\mathrm{I}}$ only to $4$th
approximation, i.e. $B_{kr}^{\mathrm{I}}$ for $r\le 4$, we have 
obtained $B_{kr}^{\mathrm{I}}$ for $r\le 7$ in this paper. 
This ensures the convergence of all the physical quantities calculated in this paper. 
It should be mentioned that our values of $B_{kr}^{\mathrm{I}}$ for
$r\le 4$ agree with Burnett's values\cite{burnett}. 
Once $B_{kr}^{\mathrm{I}}$ have been calculated,
$B_{kr}^{(1)\mathrm{I}}$ can be written down directly by replacing $\partial_{z} T$
by $\partial_{x} T$ by symmetry, owing to the properties of the
spherical harmonic function\cite{burnett}; $C_{kr}^{(1)\mathrm{I}}$ can
also be obtained similarly by replacing $\partial_{z} T$
by $\partial_{y} T$. 
Note that other terms, such as $B_{kr}^{(2)\mathrm{I}}$, do not appear
because $p$ in $B_{kr}^{(p)\mathrm{I}}$ must be $k$ or less from eq.(\ref{be14}). 
Substituting all the first-order coefficients derived here into eq.(\ref{be14}),
we can finally obtain the first-order velocity distribution function $f^{(1)}_{\mathrm{1}}$. 

\subsubsection{Solubility conditions for $\phi^{(2)}_{\mathrm{1}}$}
Since the first-order velocity distribution function $f^{(1)}_{\mathrm{1}}$ has been obtained, 
the solubility conditions of the integral equation (\ref{be5}) should be considered before we
attempt to derive an expression for $\phi^{(2)}_{\mathrm{1}}$. 
The solubility conditions for $\phi^{(2)}_{\mathrm{1}}$, that is, eqs.(\ref{be9}) lead to
the condition
\begin{eqnarray}  
\nabla \cdot {\bf J}^{(1)}=0, \label{be38}
\end{eqnarray}
where ${\bf J}^{(1)}$, i.e. the heat flux for $f^{(1)}_{\mathrm{1}}$, can be obtained as  
\begin{eqnarray}  
{\bf J}^{(1)}&\equiv&\left(\frac{2\kappa T}{m}\right)^{3}\int^{\infty}_{-\infty}\mathrm{d}{\bf c}_{\mathrm{1}}\frac{m{\bf c}_{\mathrm{1}}^{2}}{2}{\bf c}_{\mathrm{1}} f^{(1)}_{\mathrm{1}}\nonumber \\
&=&-b_{11}\frac{75}{64}\left(\frac{\kappa T}{\pi m}\right)^{\frac{1}{2}}\frac{\kappa}{\sigma^2}\nabla T, \label{be39}
\end{eqnarray} 
with the appropriate value for $b_{11}$ listed in Table \ref{b1r}. 
It must be emphasized that, since ${\bf J}^{(2)}$, i.e. the heat flux for
$f^{(2)}_{\mathrm{1}}$, does not appear, the solubility conditions of
the steady-state Boltzmann equation for $\phi^{(2)}_{\mathrm{1}}$ lead to
the heat flux being constant to second order. 
From eqs.(\ref{be38}) and (\ref{be39}), we also obtain an important relation
between $\left(\nabla T\right)^{2}$ and $\nabla^{2} T$, namely 
\begin{eqnarray}
\frac{\left(\nabla T\right)^{2}}{2T}+\nabla^{2} T=0.  
\label{be39.5}
\end{eqnarray} 
Owing to the relation (\ref{be39.5}), terms of $\nabla^{2} T$ can be replaced by terms of $\left(\nabla T\right)^{2}$. 

\subsubsection{The second order}\label{second}
We write down the results of the second-order coefficients $B_{kr}^{\mathrm{II}}$ of which $\phi^{(2)}_{\mathrm{1}}$ is composed. 
We can determine the second-order coefficients
 $B_{0r}^{\mathrm{II}}$ appearing in
eq.(\ref{be13}) as 
\begin{eqnarray}
B_{0r}^{\mathrm{II}}=\frac{b_{0r}}{\pi\sigma^{4}n^{2}T^{2}}\left(\nabla T\right)^{2}. \label{be42}
\end{eqnarray}
Values for the constants $b_{0r}$ are summarized in Table \ref{b0r}. 
The calculation of $B_{0r}^{\mathrm{II}}$ is shown in Appendix
\ref{a50}. 
It must be emphasized that Burnett did not obtain the second-order coefficients
$B_{0r}^{\mathrm{II}}$ because he was only interested in the family of the coefficient
$B_{20}^{\mathrm{II}}$
which is related to the pressure tensor $P_{ij}$.\cite{burnett}
The necessity of the second-order coefficients $B_{0r}^{\mathrm{II}}$ for 
the calculation of some physical quantities will be discussed later.  
We have calculated $B_{0r}^{\mathrm{II}}$ to $7$th approximation,
i.e. $B_{0r}^{\mathrm{II}}$ for $r\le 6$ in this paper. 

The other second-order coefficients
 $B_{kr}^{\mathrm{II}}$ in eq.(\ref{be14}) can be written in the
final form: 
\begin{eqnarray}
B_{kr}^{\mathrm{II}}=\frac{\delta_{k,2}b_{2r}}{\pi\sigma^{4}n^{2}T^{2}}
\left[2\left(\partial_{z} T\right)^{2}-\left(\partial_{x} T\right)^{2}-\left(\partial_{y} T\right)^{2}\right]. \label{be45}
\end{eqnarray}
Values for the constants $b_{2r}$ are summarized in Table \ref{b2r}. 
The calculation of $B_{kr}^{\mathrm{II}}$ is explained in Appendix
\ref{a50}. 
Our $4$th approximation $B_{20}^{\mathrm{II}}$ deviates from Burnett's
$4$th approximation $B_{20}^{\mathrm{II}}$ by a factor $1.003$. 
This deviation is considered to be due to errors in Burnett's
calculation of the second and the third terms on the right-hand side of
eq.(\ref{be123}). 
Although Burnett obtained $B_{kr}^{\mathrm{II}}$ only to $4$th approximation, i.e. $B_{kr}^{\mathrm{II}}$
for $r\le 3$, we have obtained $B_{kr}^{\mathrm{II}}$ to $7$th
approximation, i.e. $B_{kr}^{\mathrm{II}}$ for
$r\le 6$ in the present paper. 
Owing to the properties of the
spherical harmonic function\cite{burnett}, $B_{kr}^{(1)\mathrm{II}}$ can be obtained by
replacing $2(\partial_{z}
T)^{2}-(\partial_{x} T)^{2}-(\partial_{y} T)^{2}$ in eq.(\ref{be45})
by $6 \partial_{z} T \partial_{x} T$ using an axis change; similarly,
$C_{kr}^{(1)\mathrm{II}}$ can be obtained by replacing $2(\partial_{z}
T)^{2}-(\partial_{x} T)^{2}-(\partial_{y} T)^{2}$ 
by $6 \partial_{z} T \partial_{y} T$, and 
$B_{kr}^{(2)\mathrm{II}}$ can be obtained by
replacing $2(\partial_{z}
T)^{2}-(\partial_{x} T)^{2}-(\partial_{y} T)^{2}$ in eq.(\ref{be45}) by $6 [(\partial_{x}T)^{2}-(\partial_{y} T)^{2}]$ via an axis change;
$C_{kr}^{(2)\mathrm{II}}$ by replacement by $12 \partial_{x} T \partial_{y} T$. 
Note that other terms, \textit{e.g.} $B_{kr}^{(3)\mathrm{II}}$, do not appear
because $p$ in $B_{kr}^{(p)\mathrm{II}}$ must be $k$ or less, from eq.(\ref{be14}). 

One can see that both of $B_{0r}^{\mathrm{II}}$ and
$B_{kr}^{\mathrm{II}}$ are of the order of $K^2$. 
As is mentioned in Appendix \ref{a50}, we have found the fact that
$B_{kr}^{\mathrm{II}}$ for $k=4$, $6$ and $8$ do not appear, 
which strongly suggests that $B_{kr}^{\mathrm{II}}$ for all $k$ greater than $2$ do not appear. 
Thus, we expect $B_{kr}^{\mathrm{II}}=0$ for $k\ne 2$ although Burnett\cite{burnett} had
believed that they would appear. 
We have calculated the second-order coefficients now, so that we finally obtain
$f^{(2)}_{\mathrm{1}}$ by substituting the second-order coefficients
obtained here into eqs.(\ref{be13}) and (\ref{be14}).  

\subsubsection{The velocity distribution functions to second order}\label{f2x}
\begin{figure}[htbp]
\begin{center}
\includegraphics[width=15.cm]{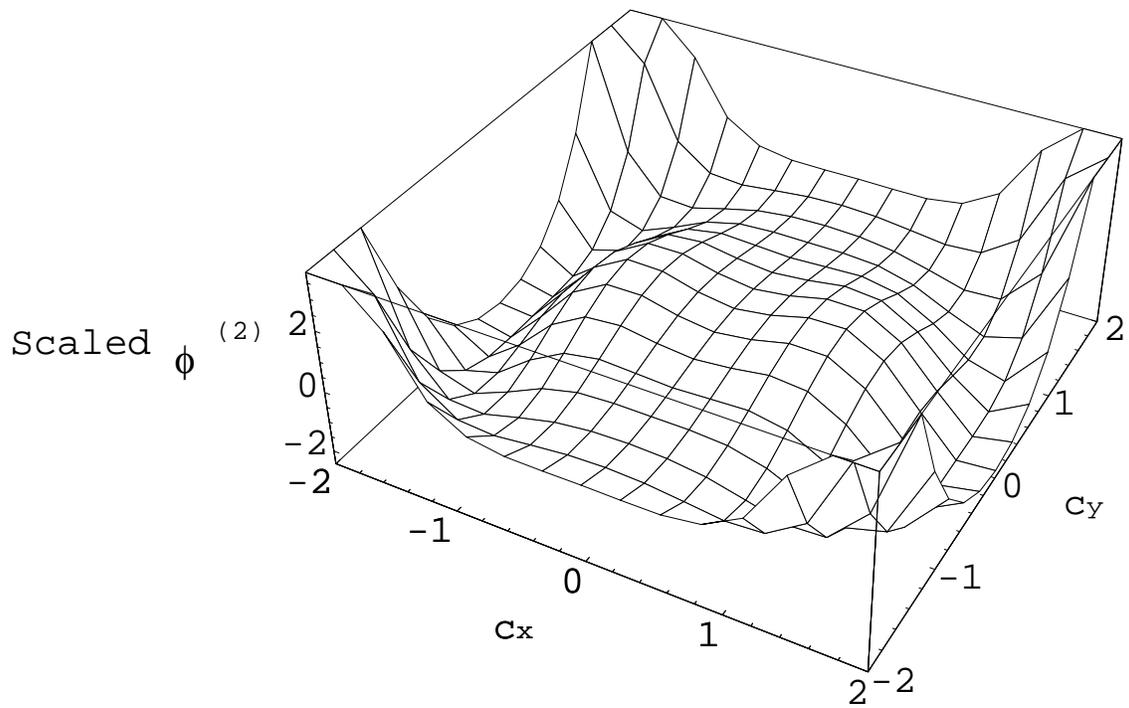} 
\caption{The scaled $\phi^{(2)}$ for hard-core
 molecules with $7$th approximation $b_{0r}$ and
 $b_{2r}$. Note that we put $c_{z}=0$. }
\label{f2cxcy}
\end{center}
\end{figure}

The velocity distribution function for hard-core molecules which we
 have derived in \S \ref{first} and \ref{second} valid to second
 order is now applied to a nonequilibrium steady-state system under the
 temperature gradient along $x$-axis. 
It can be written as  
\begin{eqnarray}  
f=f^{(0)}\{1&-&\frac{4J_{x}}{5 b_{11} n\kappa T}(\frac{m}{2\kappa T})^{\frac{1}{2}}\sum_{r\ge 1}r! b_{1r}c_{x}\Gamma(r+\frac{5}{2}) S^{r}_{\frac{3}{2}}({\bf c}^{2})\nonumber\\
&+&\frac{4096mJ_{x}^{2}}{5625b_{11}^{2}n^{2}\kappa^{3}T^{3}}[\sum_{r\ge 2}r! b_{0r}\Gamma(r+\frac{3}{2})S^{r}_{\frac{1}{2}}({\bf c}^{2})\nonumber\\
&+&\sum_{r\ge 0}r! b_{2r}(2c_{x}^{2}-c_{y}^{2}-c_{z}^{2})\Gamma(r+\frac{7}{2})S^{r}_{\frac{5}{2}}({\bf c}^{2})]
\}, \label{be46}
\end{eqnarray}
where the specific values for $b_{1r}$, $b_{0r}$ and $b_{2r}$ are found in
Tables \ref{b1r}, \ref{b0r} and \ref{b2r}, respectively,  
and $J_{x}$ corresponds to the $x$ component of
 the heat flux in eq.(\ref{be39}). 
Note that we have changed ${\bf c}_{\mathrm{1}}$ to ${\bf c}$. 
Figure \ref{f2cxcy} provides the explicit form of the second-order
velocity distribution function of the steady-state Boltzmann equation
$\phi^{(2)}({\bf c})$ for hard-core molecules with $7$th approximation $b_{0r}$ and $b_{2r}$, scaled by
$mJ_{x}^{2}/n^{2}\kappa^{3}T^{3}$. 
Figure \ref{f2cxcy} shows that the scaled $\phi^{(2)}({\bf c})$ for hard-core molecules is strained symmetrically. 

\begin{figure}[htbp]
\begin{center}
\includegraphics[width=12.cm]{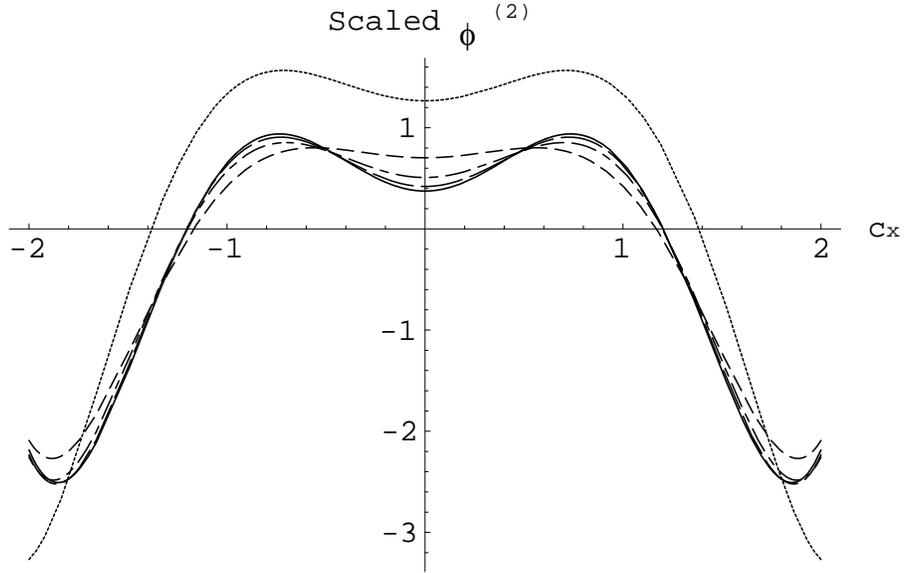} 
\caption{Comparison of the scaled $\phi^{(2)}$s for hard-core
 molecules with the scaled $\phi^{(2)}$ for Maxwell molecules. 
The dashed line, the dash-dotted line, the long-dashed line and the solid line correspond to the scaled $\phi^{(2)}$s for hard-core
 molecules with $4$th, $5$th, $6$th and $7$th approximation $b_{0r}$s and
 $b_{2r}$s, respectively. 
The dotted line is the scaled
 $\phi^{(2)}$ for Maxwell molecules. 
Note that we put $c_{y}=c_{z}=0$.}
\label{f2hikaku}
\end{center}
\end{figure}

In order to compare the macroscopic
quantities for hard-core molecules with those for Maxwell molecules, we also
adopt the velocity distribution function for
Maxwell molecules derived by Schamberg\cite{maxwell} for our calculation of the
macroscopic quantities. 
We apply Schamberg's velocity distribution function to a nonequilibrium
steady-state system under the temperature gradient along $x$-axis, that is, we
 take out the
differential coefficients with respect to time $t$ from Schamberg's
velocity distribution function and also use the solubility condition (\ref{be38})
for Maxwell molecules. 
The precise velocity distribution function of the steady-state Boltzmann
equation for
Maxwell molecules to second order finally becomes\cite{maxwell} 
\begin{eqnarray}  
f=f^{(0)}\{1&-&\frac{4J_{x}}{5 n\kappa T}(\frac{m}{2\kappa T})^{\frac{1}{2}}b_{11}c_{x}\Gamma(\frac{7}{2}) S^{1}_{\frac{3}{2}}({\bf c}^{2})
+\frac{4096mJ_{x}^{2}}{5625 n^{2}\kappa^{3}T^{3}}[\sum_{r=2, 3}r! b_{0r}\Gamma(r+\frac{3}{2})S^{r}_{\frac{1}{2}}({\bf c}^{2})\nonumber\\
&+&\sum_{r=1, 2}r! b_{2r}(2c_{x}^{2}-c_{y}^{2}-c_{z}^{2})\Gamma(r+\frac{7}{2})S^{r}_{\frac{5}{2}}({\bf c}^{2})]
\}.   
\label{be465}
\end{eqnarray}
As can be seen from eqs.(\ref{be46}) and (\ref{be465}), 
the explicit form of the velocity distribution function for hard-core molecules becomes
the sum of an infinite series of Sonine
polynomials, while the precise form of the velocity
distribution function for Maxwell molecules is the sum of five Sonine
polynomials. 
Figure \ref{f2hikaku} gives the direct comparison of the scaled
$\phi^{(2)}$s in eq.(\ref{be46}) with $4$th, $5$th, $6$th and $7$th approximation $b_{0r}$s and $b_{2r}$s with the scaled
$\phi^{(2)}$ in eq.(\ref{be465}). 
It should be mentioned that, as Fig.\ref{f2hikaku} shows, the scaled
$\phi^{(2)}$ in eq.(\ref{be46}) has not yet converged to $4$th
approximation. 
Furthermore, in this paper, we adopt the precise expression of the corresponding velocity distribution
function for the steady-state
BGK equation, though we do not write down the explicit form of it, which can be reduced from the general form of the the
Chapman-Enskog solution for the steady-state BGK equation to arbitrary
order\cite{santos3,santos31,santos4,santos41,santos42}. 
We mention that $\phi^{(1)}$ for the steady-state BGK equation 
 is identical to that for the steady-state Boltzmann equation for
Maxwell molecules appearing in eq.(\ref{be465}), while $\phi^{(2)}$s are different from each other. 

\section{Thermodynamic Quantities}\label{application}
We can introduce the general form of the heat flux as 
\begin{eqnarray}  
J_{x}=-\varpi T^{\varphi} \partial_{x} T,    
\label{be470}
\end{eqnarray}
where $\varphi$ indicates temperature dependence of the thermal
conductivity and $\varpi$ is a constant that depends upon microscopic
models. 
$J_{x}$ is constant from the solubility condition (\ref{be38}).  
For example, 
$\varphi$ is calculated as $1/2$ for hard-core molecules and $1$ for Maxwell
molecules; 
$\varpi$ is determined as 
 $75 b_{11}\kappa(\kappa /\pi m)^{1/2}/64\sigma^2$  with $b_{11}\simeq
 1.025$ for hard-core molecules ( see eq.(\ref{be39})) and  
$5 \kappa^{2}(2m/G)^{1/2}/4A$ for 
Maxwell molecules\cite{maxwell1}, where $G$ is the constant of proportionality between
the intermolecular force and the reciprocal fifth power of the distance,
and $A$ is a number constant erroneously evaluated as $1.3682$ by
Maxwell\cite{maxwell0,maxwell01} and recalculated as $1.3700$ by
Chapman\cite{recalculated}. 
Note that $\varphi$ and $\varpi$ cannot be determined explicitly from the BGK
equation. 
From eq.(\ref{be470}), the temperature profile $T(x)$ in the
nonequilibrium steady state can be determined as
\begin{eqnarray}  
T(x)
&=&[T(0)^{\varphi+1}-(\varphi+1)\frac{J_{x}}{\varpi}x]^{\frac{1}{\varphi+1}}\nonumber \\
&\simeq&  T(0)[1-\frac{J_{x}}{\varpi T(0)^{\varphi+1}}x-\frac{\varphi J_{x}^{2}}{2\varpi^{2} T(0)^{2\varphi+2}}x^{2}],
\label{be47}
\end{eqnarray}
to second order. 
The temperature profile $T(x)$ becomes nonlinear except for
$\varphi=0$. 
This fact is in accord with a numerical result that a temperature profile
becomes nonlinear as the heat flux becomes
 larger.\cite{lorentz} 

Using eq.(\ref{be17.5}), the pressure tensor $P_{ij}$ in the nonequilibrium steady
state can be obtained as
\begin{eqnarray}  
P_{ij}&=& n\kappa T[\delta_{ij}+\lambda_{P}^{ij}\frac{mJ_{x}^{2}}{n^{2}\kappa^{3}T^{3}}],\label{be51}
\end{eqnarray}
with the unit tensor $\delta_{ij}$ and the tensor components
$\lambda_{P}^{ij}$ given in Table \ref{macro1}. 
Note that the off-diagonal components of $\lambda_{P}^{ij}$
are zero and $\lambda_{P}^{yy}=\lambda_{P}^{zz}$ is satisfied. 
The values of $\lambda_{P}^{ij}$ for $7$th approximation $b_{1r}$,
$b_{0r}$ and $b_{2r}$ for hard-core molecules, i.e. $7$th approximation $\lambda_{P}^{ij}$
seems to be converged to three significant figures, as can be seen from Table \ref{macro1}. 
We find that $\lambda_{P}^{ij}$ for hard-core molecules differs from
that for Maxwell molecules not only quantitatively
  but also qualitatively:  $\lambda_{P}^{xx}\ne
  \lambda_{P}^{yy}=\lambda_{P}^{zz}$ for hard-core molecules in the
  nonequilibrium steady state, while $\lambda_{P}^{xx}=\lambda_{P}^{yy}=\lambda_{P}^{zz}=0$ for Maxwell molecules. 
For hard-core molecules, $P_{xx}$ becomes smaller than $P_{yy}$ and
$P_{zz}$ regardless of the sign of $J_{x}$. 
It is also important that, since $n\kappa T$ is required to be uniform from
the solubility conditions for $\phi^{(1)}$ in eq.(\ref{be7}), we find
from eq.(\ref{be47}) that $P_{ij}$ in eq.(\ref{be51}) is uniform to
second order in the steady state. 
Additionally, since the
temperature profile $T(x)$ has already been given in eq.(\ref{be47}),
we can determine the density profile $n(x)$ to second order for uniform $P_{ij}$. 
Note that the equation of state in the nonequilibrium steady
state is not modified to first order, and that $\lambda_{P}^{ij}$ for
the steady-state BGK equation is identical with that for Maxwell molecules. 

Each component of the kinetic temperature in the nonequilibrium steady
state, i.e. $T_{\mathrm{i}}$ for ${\mathrm{i}}=x,y$ and $z$ is calculated as
\begin{eqnarray}  
\frac{n\kappa T_{\mathrm{i}}}{2}&\equiv&\left(\frac{2\kappa T}{m}\right)^{\frac{5}{2}}\int^{\infty}_{-\infty}\mathrm{d}{\bf c}\frac{mc_{\mathrm{i}}^{2}}{2}f \nonumber \\
&=&\frac{n\kappa T}{2}[1+\lambda_{T_{\mathrm{i}}}\frac{mJ_{x}^{2}}{n^{2}\kappa^{3}T^{3}}],\label{be54}
\end{eqnarray} 
for ${\mathrm{i}}=x, y$ and $z$. 
Values for the constants $\lambda_{T_{\mathrm{i}}}$ for ${\mathrm{i}}=x,y$ and $z$ are given in
Table \ref{macro1}. 
Note the relation $\lambda_{T_{y}}=\lambda_{T_{z}}$. 
$\lambda_{T_{\mathrm{i}}}$ with $7$th approximation $b_{1r}$, $b_{0r}$ and $b_{2r}$ for hard-core
molecules, i.e. $7$th approximation $\lambda_{T_{\mathrm{i}}}$
seems to converge to three significant figures, according to the results
reported in Table \ref{macro1}. 
Since $n\kappa T$ and $J_{x}$ are uniform, we see that the correction
term of $T_{\mathrm{i}}$ in
eq.(\ref{be54}) is independent of $x$. 
The value of $\lambda_{T_{\mathrm{i}}}$ for hard-core molecules is seen to be 
qualitatively different from that for Maxwell molecules: 
  $\lambda_{T_{x}}\ne\lambda_{T_{y}}=\lambda_{T_{z}}$ for hard-core
  molecules in the nonequilibrium steady-state, while 
$\lambda_{T_{x}}=\lambda_{T_{y}}=\lambda_{T_{z}}=0$ for Maxwell
molecules in that state. 
For hard-core molecules, $T_{x}$ becomes smaller than $T_{y}$ and
$T_{z}$ regardless of the sign of $J_{x}$, which means that the
motion of hard-core molecules along the heat flux becomes dull. 
We note that $T_{\mathrm{i}}$ for
the steady-state BGK equation is identical with that for Maxwell
molecules, and that, to first order, $T_{\mathrm{i}}$ for hard-core molecules is isotropic. 

The Shannon entropy in the nonequilibrium steady
state $S$ is defined via
\begin{eqnarray}  
S&\equiv&-\kappa\left(\frac{2\kappa T}{m}\right)^{\frac{3}{2}}\int^{\infty}_{-\infty}\mathrm{d}{\bf c}f\log f \nonumber \\
&=&-n\kappa\log \left[n\left(\frac{m}{2\pi \kappa T}\right)^{\frac{3}{2}}\right]+\frac{3}{2}n\kappa+\lambda_{S}\frac{mJ_{x}^{2}}{n\kappa^{2}T^{3}}.\label{be56}
\end{eqnarray}
Values for the constant $\lambda_{S}$ are given in Table \ref{macro1}: 
$\lambda_{S}$ for $7$th approximation $b_{1r}$, $b_{0r}$ and $b_{2r}$ for hard-core
molecules, i.e. $7$th approximation $\lambda_{S}$ 
seems to converge to four significant figures, see Table \ref{macro1}. 
It is found that $\lambda_{S}$ for hard-core molecules is close to that for
Maxwell molecules, while the latter is identical with that for the steady-state BGK equation. 
Note that the Shannon entropy in the nonequilibrium steady
state is not modified to first order. 

\section{Discussion}\label{discussion}
It is noteworthy that all macroscopic quantities now become able to be
calculated directly from the explicit velocity distribution
functions of the steady-state Boltzmann equation to second order. 
Actually, as in Appendix \ref{a6}, we examine the
nonequilibrium steady-state thermodynamics(SST) proposed by Oono
and Paniconi\cite{oono}, and extended by Sasa
and Tasaki\cite{sst} by using the velocity distribution function to
second order for hard-core molecules we have
derived in \S \ref{f2} and that for Maxwell molecules. 
As a result, our results do not support SST. 
We mention that the full information of the second-order coefficients
$B_{0r}^{\mathrm{II}}$ and $B_{kr}^{\mathrm{II}}$ is necessary for the calculation of
some physical quantities in the test of SST. (see Appendix \ref{a6})  
Furthermore, as indicated in the introduction, the other important application of the
explicit second-order velocity distribution function for hard-core molecules is to evaluate the nonequilibrium effect on
rate of chemical reaction of gases. 
The line-of-centers model\cite{present} which is accepted as a standard one requires
the full form of the second-order velocity distribution function for
hard-core molecules\cite{fort}. 
Our result enables us to calculate the reaction rate.\cite{kim1} 

We have found that there are qualitative differences between hard-core molecules and Maxwell molecules in the nonequilibrium steady state:
 second-order corrections appear for hard-core molecules in the pressure tensor $P_{ij}$
and the kinetic temperature $T_{\mathrm{i}}$, while no correction to these
 quantities appears for Maxwell molecules, as
 Table \ref{macro1} shows. 
It should be noted that the qualitative differences between hard-core molecules
 and Maxwell molecules still appear no matter which boundary condition is adopted, that is, the isotropy and  
the anisotropy of the pressure tensor in eq.(\ref{be51})
and the kinetic temperature in eq.(\ref{be54}) are not affected by any
 kinds of boundary conditions. 
It is conjectured that these differences are attributed to the special nature of Maxwell
 molecules: $\Delta_{kr}$ for Maxwell molecules, i.e. $\Delta_{kr}^{\mathrm{M}}$ is independent of
the magnitude of the relative velocity $g$, because $F_{kr}^{0}(\chi)$
 is independent of $g$. 
In general, $\Delta_{kr}$ for molecules which interact with each
other by a central force depends on $g$, because $F_{kr}^{\mu}(\chi)$ generally depends on $g$. 
Therefore, it may be suggested that molecules which interact via a
central potential still have the qualitative differences from Maxwell
molecules: second-order corrections in pressure tensor $P_{ij}$
and kinetic temperature $T_{\mathrm{i}}$ may appear also for such molecules as
well as hard-core molecules. 

It is also found that the pressure tensor $P_{ij}$
and the kinetic temperature $T_{\mathrm{i}}$ for the steady-state BGK equation are qualitatively
different from those for the steady-state
Boltzmann equation for hard-core molecules but
agree with those for the steady-state
Boltzmann equation for Maxwell molecules, as illustrated in Table
\ref{macro1}. 
Since the Chapman-Enskog solution of the
steady-state BGK equation is asymptotically
correct\cite{santos3,santos31,santos4,santos41,santos42}, we may conclude that the steady-state BGK
equation does not capture the essence of hard-core molecules, but
captures that of Maxwell-type molecules. 
This conclusion indicates the possibility that even the exact solution of the
steady-state BGK equation can be applied only to
Maxwell-type molecules. 

Finally, we consider the possibility of the existence of a universal velocity distribution
function in the nonequilibrium steady state. 
When we express the velocity distribution function for
 hard-core molecules to second order using the heat flux $J_{x}$ as in eq.(\ref{be46}), it becomes independent of the
 diameter $\sigma$ of the hard-core molecules. 
However, the explicit form of the velocity distribution function for
 hard-core molecules (\ref{be46}) definitely differs from the precise
 form of the velocity distribution function for Maxwell molecules
 (\ref{be465}) as Fig.\ref{f2hikaku} shows. 
Actually, as mentioned above, we have shown that the results calculated from the former are
qualitatively different from those calculated from the latter. 
These results indicate that the
characteristics of microscopic models appear in a 
nonequilibrium steady state solution and affect even qualitatively 
the macroscopic quantities of a system in that nonequilibrium steady
state. 
It can be concluded that a universal velocity distribution
function does not exist in the nonequilibrium steady state even if the velocity distribution
function is expressed only in terms of macroscopic quantities.\cite{kim} 

\section{Conclusion}\label{conclusion}
The velocity distribution function of the steady-state Boltzmann equation for
 hard-core molecules subject to a temperature gradient has been derived
 explicitly to second order in the density and the temperature gradient
 as shown explicitly in eq.(\ref{be46}) and Fig.\ref{f2cxcy}. 
In the nonequilibrium steady-state system, qualitative differences between hard-core molecules
 and Maxwell molecules are found in the pressure tensor (\ref{be51})
 and the kinetic temperature (\ref{be54}): it appears that Maxwell
 molecules do not possess the characteristics of other models of molecules which interact with each
other by central forces in the nonequilibrium
 steady-state system. 
Additionally, we have found that the steady-state BGK equation belongs to the same
 universality class as Maxwell molecules, and that it does not capture
 the essence of hard-core molecules.  
We finally conclude that
 a universal velocity distribution
function does not seem to exist, as Fig.\ref{f2hikaku}
explicitly shows, even when the velocity distribution
function is expressed only in terms of macroscopic quantities. 

\section*{Acknowledgments}
We would like to express our sincere gratitude to S. Sasa who has made us aware of the significance
of understanding nonequilibrium steady-state phenomena.  
This research was essentially inspired by him. 
We thank H. Tasaki who has always made the crucial, interesting and cheerful discussions with
us and who has encouraged us to carry out these calculations. 
The authors also appreciate Ooshida. T, A. Yoshimori, 
M. Sano, J. Wakou, K. Sato, H. Kuninaka, T. Mizuguti, T. Chawanya, S. Takesue 
and H. Tomita for fruitful discussions and useful comments. 
We thank A. Santos to provide helpful and important papers for us. 
This study has been supported partially by the Hosokawa Powder Technology
Foundation, the Inamori Foundation and Grant-in-Aid for Scientific Research (No. 13308021).  

\appendix
\section{The Calculation of $\Omega_{\lowercase{kr}}$}\label{a3}
From the definition of $Q_{kr}$, $\Omega_{kr}$ can be
calculated using the mathematical properties of the spherical harmonic
functions and the Sonine polynomials.\cite{burnett,maxwell} 
One can calculate $D_{k,r}$, $E_{k,r}$ and $G_{k,r}$ defined as   
\begin{eqnarray}  
D_{k,r}\equiv \frac{1}{n}<c_{{\mathrm{1}} x} Q_{kr}>_{av}, \quad E_{k,r}\equiv\frac{1}{n}<c_{{\mathrm{1}} y} Q_{kr}>_{av} \quad \mathrm{and} \quad G_{k,r}\equiv \frac{1}{n}<c_{{\mathrm{1}} z} Q_{kr}>_{av}. 
\end{eqnarray}
The results can be written as 
\begin{eqnarray}
D_{k,r}&=&\frac{1}{2k+3}[(k+r+\frac{3}{2})B_{k+1,r}^{(1)}-B_{k+1,r-1}^{(1)}]-\frac{1}{2k-1}[B_{k-1,r}^{(1)}-(r+1)B_{k-1,r+1}^{(1)}],  
\label{be22}
\end{eqnarray}
and 
\begin{eqnarray}  
E_{k,r}&=&\frac{1}{2k+3}[(k+r+\frac{3}{2})C_{k+1,r}^{(1)}-C_{k+1,r-1}^{(1)}]-\frac{1}{2k-1}[C_{k-1,r}^{(1)}-(r+1)C_{k-1,r+1}^{(1)}],\label{be23}
\end{eqnarray}
and
\begin{eqnarray}  
G_{k,r}&=&\frac{k+1}{2k+3}[(k+r+\frac{3}{2})B_{k+1,r}-B_{k+1,r-1}]-\frac{k}{2k-1}[B_{k-1,r}-(r+1)B_{k-1,r+1}]. 
\label{be24}
\end{eqnarray}
One can also obtain 
\begin{eqnarray}  
<c_{{\mathrm{1}} x} \partial_{x} Q_{kr}>_{av}=-n\frac{\partial_{x} T}{T}[(r+\frac{k}{2})D_{k,r}-D_{k,r-1}].\label{be25}
\end{eqnarray}
Similarly 
$<c_{{\mathrm{1}} y} \partial_{y} Q_{kr}>_{av}$ 
and $<c_{{\mathrm{1}} z} \partial_{z} Q_{kr}>_{av}$ are obtained by replacing the
differential coefficients with respect to $x$ by the corresponding 
differential coefficients with respect to $y$ and
$z$, the $D_{k,r}$'s by the
corresponding $E_{k,r}$'s and $G_{k,r}$'s, respectively.  
Substituting these results into eq.(\ref{be19.0}), $\Omega_{kr}$ finally
becomes eq.(\ref{be25.5}). 

\section{The Details of $F_{kr}^{1}(\chi)$}\label{a2}
The details of $F_{kr}^{1}(\chi)$ are as follows.  
Substituting the general forms of $f_{\mathrm{1}}$, $f_{\mathrm{2}}$ in eq.(\ref{be13}) 
and $Q^{\prime}_{kr}$ in eq.(\ref{be20}) into $F_{kr}^{1}(\chi)$ in
eq.(\ref{be27}), $F_{kr}^{1}(\chi)$ can be written as\cite{burnett}
\begin{eqnarray}
F_{kr}^{1}(\chi)=\sum_{n_{\mathrm{1}}, n_{\mathrm{2}}, k_{\mathrm{1}}, k_{\mathrm{2}}, k_{\mathrm{1}} \ge p_{\mathrm{1}}\ge 0, k_{\mathrm{2}} \ge p_{\mathrm{2}}\ge 0}W_{k,k_{\mathrm{1}},k_{\mathrm{2}}}^{n_{\mathrm{1}}, n_{\mathrm{2}},p_{\mathrm{1}},p_{\mathrm{2}}}\Xi_{k,k_{\mathrm{1}},k_{\mathrm{2}}}^{r,n_{\mathrm{1}},n_{\mathrm{2}},p_{\mathrm{1}},p_{\mathrm{2}}}(\chi),\label{be34.5}
\end{eqnarray}
where the summation with respect to $p_{\mathrm{1}}$ and $p_{\mathrm{2}}$ is performed from $0$
to $k_{\mathrm{1}}$ and $k_{\mathrm{2}}$, respectively as seen in eq.(\ref{be14}).  
Here $\Xi_{k,k_{\mathrm{1}},k_{\mathrm{2}}}^{r,n_{\mathrm{1}},n_{\mathrm{2}},p_{\mathrm{1}},p_{\mathrm{2}}}(\chi)$
is the characteristic integral defined as 
\begin{eqnarray}
\Xi_{k,k_{\mathrm{1}},k_{\mathrm{2}}}^{r,n_{\mathrm{1}},n_{\mathrm{2}},p_{\mathrm{1}},p_{\mathrm{2}}}(\chi)&\equiv& \Gamma(k_{\mathrm{1}}+n_{\mathrm{1}}+\frac{3}{2})\Gamma(k_{\mathrm{2}}+n_{\mathrm{2}}+\frac{3}{2}) \left(\frac{2\kappa T}{m}\right)^{3}\int \int \int Y_{k}({\bf c}_{\mathrm{1}}^{\prime}) Y^{(p_{\mathrm{1}})}_{k_{\mathrm{1}}}({\bf c}_{\mathrm{1}})Y^{(p_{\mathrm{2}})}_{k_{\mathrm{2}}}({\bf c}_{\mathrm{2}})\nonumber\\
&\times& \exp[-(c_{\mathrm{1}}^{2}+c^{2}_{\mathrm{2}})] S^{r}_{k+\frac{1}{2}}({\bf c}_{\mathrm{1}}^{\prime 2})S^{n_{\mathrm{1}}}_{k_{\mathrm{1}}+\frac{1}{2}}({\bf c}_{\mathrm{1}}^{2})S^{n_{\mathrm{2}}}_{k_{\mathrm{2}}+\frac{1}{2}}({\bf c}_{\mathrm{2}}^{2}) g \mathrm{d}\epsilon \mathrm{d}{\bf c}_{\mathrm{2}}\mathrm{d}{\bf c}_{\mathrm{1}}.\label{be29}
\end{eqnarray}
Note that the integrals containing products like
$Y^{(p_{\mathrm{1}})}_{k_{\mathrm{1}}}({\bf
c}_{\mathrm{1}})Z^{(p_{\mathrm{2}})}_{k_{\mathrm{2}}}({\bf c}_{\mathrm{2}})$ are
zero, owing to the orthogonality properties of the spherical harmonic functions, while those containing $Z^{(p_{\mathrm{1}})}_{k_{\mathrm{1}}}({\bf
c}_{\mathrm{1}})Z^{(p_{\mathrm{2}})}_{k_{\mathrm{2}}}({\bf c}_{\mathrm{2}})$ are
identical with the corresponding integrals
$\Xi_{k,k_{\mathrm{1}},k_{\mathrm{2}}}^{r,n_{\mathrm{1}},n_{\mathrm{2}},p_{\mathrm{1}},p_{\mathrm{2}}}(\chi)$. 
The factor $W_{k,k_{\mathrm{1}},k_{\mathrm{2}}}^{n_{\mathrm{1}},
n_{\mathrm{2}},p_{\mathrm{1}},p_{\mathrm{2}}}$ in eq.(\ref{be34.5}) is defined as
\begin{eqnarray}
W_{k,k_{\mathrm{1}},k_{\mathrm{2}}}^{n_{\mathrm{1}}, n_{\mathrm{2}},p_{\mathrm{1}},p_{\mathrm{2}}}\equiv n^{2}\left(\frac{m}{2\pi\kappa T}\right)^{3}\left(\frac{m}{2\kappa T}\right)^\frac{k+k_{\mathrm{1}}+k_{\mathrm{2}}}{2}n_{\mathrm{1}}! n_{\mathrm{2}}! (k+\frac{1}{2}) \sqrt{\pi}\nonumber \\
 \Psi_{k_{\mathrm{1}},k_{\mathrm{2}}}^{p_{\mathrm{1}},p_{\mathrm{2}}} (B_{k_{\mathrm{1}}n_{\mathrm{1}}}^{(p_{\mathrm{1}})}B_{k_{\mathrm{2}}n_{\mathrm{2}}}^{(p_{\mathrm{2}})}+C_{k_{\mathrm{1}}n_{\mathrm{1}}}^{(p_{\mathrm{1}})}C_{k_{\mathrm{2}}n_{\mathrm{2}}}^{(p_{\mathrm{2}})}), \label{be34}
\end{eqnarray}
which is obtained from the prefactors and the coefficients in the general form of
$f_{\mathrm{1}}$, $f_{\mathrm{2}}$ in eq.(\ref{be13}) and
$Q^{\prime}_{kr}$ in eq.(\ref{be20}). 
Note that $C_{k_{\mathrm{1}}n_{\mathrm{1}}}^{(p_{\mathrm{1}})}C_{k_{\mathrm{2}}n_{\mathrm{2}}}^{(p_{\mathrm{2}})}$
appear from the integrals containing $Z^{(p_{\mathrm{1}})}_{k_{\mathrm{1}}}({\bf
c}_{\mathrm{1}})Z^{(p_{\mathrm{2}})}_{k_{\mathrm{2}}}({\bf
c}_{\mathrm{2}})$, and that $B_{kr}^{(0)}=B_{kr}$ and $C^{(0)}_{kr}=0$ from
eqs.(\ref{be14}), (\ref{be15}), (\ref{be16}) and (\ref{be17}).  
The constant
$\Psi_{k_{\mathrm{1}},k_{\mathrm{2}}}^{p_{\mathrm{1}},p_{\mathrm{2}}}$
is defined as 
\begin{eqnarray}
\Psi_{k_{\mathrm{1}},k_{\mathrm{2}}}^{p_{\mathrm{1}},p_{\mathrm{2}}}=1  \quad \mathrm{for} \quad p_{\mathrm{1}}=p_{\mathrm{2}}=0,
\label{be34.7}
\end{eqnarray}
and
\begin{eqnarray}
\Psi_{k_{\mathrm{1}},k_{\mathrm{2}}}^{p_{\mathrm{1}},p_{\mathrm{2}}}
=\frac{4(k_{\mathrm{1}}-p_{\mathrm{1}})!(k_{\mathrm{2}}-p_{\mathrm{1}})!}{(k_{\mathrm{1}}+p_{\mathrm{1}})!(k_{\mathrm{2}}+p_{\mathrm{1}})!} \quad \mathrm{for} \quad p_{\mathrm{1}}=p_{\mathrm{2}}\ne 0. 
\label{be34.8}
\end{eqnarray}
Here $\Psi_{k_{\mathrm{1}},k_{\mathrm{2}}}^{p_{\mathrm{1}},p_{\mathrm{2}}}$
for $p_{\mathrm{1}}\ne p_{\mathrm{2}}$
is not necessary for our calculation.\cite{burnett,kim2} 
It is found that we need only to evaluate the characteristic integral
$\Xi_{k,k_{\mathrm{1}},k_{\mathrm{2}}}^{r,n_{\mathrm{1}},n_{\mathrm{2}},p_{\mathrm{1}},p_{\mathrm{1}}}(\chi)$
in order to calculate $F_{kr}^{1}(\chi)$. 
Our calculation of
$\Xi_{k,k_{\mathrm{1}},k_{\mathrm{2}}}^{r,n_{\mathrm{1}},n_{\mathrm{2}},p_{\mathrm{1}},p_{\mathrm{1}}}(\chi)$
is written in ref.\cite{kim2}. 
Our calculation has been performed mainly based on Burnett's
method\cite{burnett}.  
We have, however, made some modifications on his method, which make the
calculation of
$\Xi_{k,k_{\mathrm{1}},k_{\mathrm{2}}}^{r,n_{\mathrm{1}},n_{\mathrm{2}},p_{\mathrm{1}},p_{\mathrm{1}}}(\chi)$ 
much easier.\cite{kim2} 
We emphasize that our calculation could not be carried out completely if 
we did not make the modifications on Burnett's method. 

Once the characteristic integral $\Xi_{k,
k_{\mathrm{1}},k_{\mathrm{2}}}^{r,n_{\mathrm{1}},
n_{\mathrm{2}},p_{\mathrm{1}},p_{\mathrm{1}}}(\chi)$ has been derived,
$F_{kr}^{1}(\chi)$ is now calculated from eq.(\ref{be34.5}) with
$W_{k,k_{\mathrm{1}},k_{\mathrm{2}}}^{n_{\mathrm{1}},
n_{\mathrm{2}},p_{\mathrm{1}},p_{\mathrm{1}}}$ in eq.(\ref{be34}). 
Note that the coefficient term $B_{k_{\mathrm{1}}n_{\mathrm{1}}}^{(p_{\mathrm{1}})}B_{k_{\mathrm{2}}n_{\mathrm{2}}}^{(p_{\mathrm{2}})}+C_{k_{\mathrm{1}}n_{\mathrm{1}}}^{(p_{\mathrm{1}})}C_{k_{\mathrm{2}}n_{\mathrm{2}}}^{(p_{\mathrm{2}})}$ in $W_{k,k_{\mathrm{1}},k_{\mathrm{2}}}^{n_{\mathrm{1}},
n_{\mathrm{2}},p_{\mathrm{1}},p_{\mathrm{1}}}$ can be determined for
each order when the suffices $k_{\mathrm{1}}$, $k_{\mathrm{2}}$,
$n_{\mathrm{1}}$, $n_{\mathrm{2}}$, $p_{\mathrm{1}}$ and $p_{\mathrm{2}}$ are specified. 

\section{The Calculation of the First Order Coefficients $B_{kr}^{\mathrm{I}}$}\label{a10}
Let us explain how to obtain the first-order coefficients, that is, 
how to solve the integral equation (\ref{be4}). 
To begin with, we calculate $\Omega_{kr}^{\mathrm{H}}$ in
eq.(\ref{be25.5}) to first order; 
$\Omega_{kr}^{\mathrm{H}}$ for first order corresponds to the right-hand side
of eq.(\ref{be4}). 
For first order, $\Omega^{\mathrm{H}}_{kr}$ in eq.(\ref{be25.5}) can be calculated by substituting
$B_{00}=1$ into the expressions of $D_{k,r}$, $E_{k,r}$
and $G_{k,r}$ in eqs.(\ref{be22}), (\ref{be23}) and (\ref{be24}): the coefficient $B_{00}=1$
corresponds to $f_{\mathrm{1}}=f^{(0)}_{\mathrm{1}}$, and higher-order terms do not
appear in $\Omega^{\mathrm{H}}_{kr}$ for first order. 
$\Omega_{kr}^{\mathrm{H}}$ for first order finally becomes 
\begin{eqnarray}
\Omega^{\mathrm{H}}_{1r}=-\frac{n}{T}\left(\frac{m}{2\kappa T}\right)^{-\frac{1}{2}}\frac{\partial T}{\partial z}\delta_{1,r}. 
\label{be36}
\end{eqnarray} 

Now $\Omega^{\mathrm{H}}_{kr}$ for first order is found to vanish unless
$k=1$, so that we need calculate only $\Delta^{\mathrm{H}}_{1,r}$ for first order; as was
mentioned in the end of \S \ref{solve}, we do not need to consider the case in
which the right-hand side of eq.(\ref{be4}) becomes zero.\cite{resi} 
To derive $\Delta^{\mathrm{H}}_{1,r}$ in eq.(\ref{be28}) for first order, 
we must calculate both $W_{1,k_{\mathrm{1}},k_{\mathrm{2}}}^{n_{\mathrm{1}},
n_{\mathrm{2}},p_{\mathrm{1}},p_{\mathrm{1}}}$ and
$\Xi_{1,k_{\mathrm{1}},k_{\mathrm{2}}}^{r,n_{\mathrm{1}},n_{\mathrm{2}},p_{\mathrm{1}},p_{\mathrm{1}}}$ 
in $F_{1,r}^{1}(\chi)$ of eq.(\ref{be34.5}) for first order as was shown
in the previous subsection.    
The result for $\Delta^{\mathrm{H}}_{1,r}$ to first order can be written
finally in the form 
\begin{eqnarray}
\Delta^{\mathrm{H}}_{1r}=B_{00}\sum_{n_{\mathrm{1}}} B_{1n_{\mathrm{1}}}^{\mathrm{I}} M_{1,1,0}^{r,n_{\mathrm{1}},0,0,0}, \label{be35}
\end{eqnarray}
where the set of the coefficients $B_{1n_{\mathrm{1}}}^{\mathrm{I}}B_{00}$ is obtained from 
$W_{1,1,0}^{n_{\mathrm{1}},0,0,0}$ in eq.(\ref{be34}). 
$f_{\mathrm{1}}$ in eq.(\ref{be27}) contains  only $B_{00}=1$ and
the first-order coefficients, i.e. the family of $B_{k_{\mathrm{1}}n_{\mathrm{1}}}^{\mathrm{I}}$ to first order; $f_{\mathrm{2}}$ in
eq.(\ref{be27}) also contains $B_{00}=1$ and the family of
$B_{k_{\mathrm{2}}n_{\mathrm{2}}}^{\mathrm{I}}$ to first order. 
Thus, we obtain only the term $B_{1n_{\mathrm{1}}}^{\mathrm{I}}B_{00}$ from $W_{1,1,0}^{n_{\mathrm{1}},0,0,0}$ to first order using the
fact that $F_{kr}^{1}(\chi)=0$ unless
$k=|k_{\mathrm{1}}-k_{\mathrm{2}}|+2q$. 
Note that it is sufficient to consider only the case for
$k_{\mathrm{1}}\ge k_{\mathrm{2}}$ as is explained in
refs.\cite{burnett,kim2}, and that we set $q=0$. 
The matrix $M_{1,1,0}^{r,n_{\mathrm{1}},0,0,0}$ is thus obtained 
\begin{eqnarray}
M_{1,1,0}^{r,n_{\mathrm{1}},
0,0,0}&=&\frac{3n^{2}\sigma^{2}m^{4}n_{\mathrm{1}}!}{64 \pi^{\frac{5}{2}}\kappa^{4} T^{4}} \nonumber\\
&\times& \int_{0}^{\pi} [\Xi_{1,1,0}^{r,n_{\mathrm{1}},0,0,0}(\chi)-\Xi_{1,1,0}^{r,n_{\mathrm{1}},0,0,0}(0) ]\sin\frac{\chi}{2} \cos\frac{\chi}{2} \mathrm{d}\chi, \label{be28.1}
\end{eqnarray}
using eqs.(\ref{be28}), (\ref{be34.5}) and (\ref{be34}). 

For $k=1$, eq.(\ref{be0}) gives simultaneous equations determining
the first-order coefficients $B_{1n_{\mathrm{1}}}^{\mathrm{I}}$, i.e. 
\begin{eqnarray}
\Omega^{\mathrm{H}}_{1r}=\sum_{n_{\mathrm{1}}\ge 1} B_{1n_{\mathrm{1}}}^{\mathrm{I}}M_{1,1,0}^{r,n_{\mathrm{1}},0,0,0},\label{be0.1}
\end{eqnarray}
from eqs.(\ref{be36}) and (\ref{be35}). 
Note that we need only to obtain the first-order coefficients
$B_{1n_{\mathrm{1}}}^{\mathrm{I}}$ for $n_{\mathrm{1}}\ge 1$, because $B_{10}=0$ from
eq.(\ref{be18}). 
We have calculated the matrix $M_{1,1,0}^{r,n_{\mathrm{1}},
0,0,0}$ for $1 \le r \le 7$ and $1 \le n_{\mathrm{1}} \le 7$   from
eq.(\ref{be28.1}), and we have confirmed that $M_{1,1,0}^{0,n_{\mathrm{1}},
0,0,0}$ for $1\le n_{\mathrm{1}}\le 7$ calculated from
eq.(\ref{be28.1}) vanishes.  
Our result for $M_{1,1,0}^{r,n_{\mathrm{1}},
0,0,0}$ for $1 \le r \le 7$ and $1 \le n_{\mathrm{1}} \le 7$ is given in ref.\cite{address}. 
At last, we can determine the first-order coefficients
$B_{1n_{\mathrm{1}}}^{\mathrm{I}}$ by solving the simultaneous
equations (\ref{be0.1}), that is, $B_{1n_{\mathrm{1}}}^{\mathrm{I}}$ can be obtained as
\begin{eqnarray}
B_{1n_{\mathrm{1}}}^{\mathrm{I}}=\sum_{r\ge 1} \Omega^{\mathrm{H}}_{1r} (M_{1,1,0}^{r,n_{\mathrm{1}},0,0,0})^{-1}, \label{be35.5}
\end{eqnarray}
where $X^{-1}$ represents the inverse matrix of a matrix $X$. 
Note that we have confirmed, using eqs.(\ref{be36}) and (\ref{be28.1}), that both
sides of eq.(\ref{be0.1}) for $r=0$ vanish, so that we need only calculate 
both sides of eq.(\ref{be0.1}) for $r\ge 1$. 
Finally, the results of the first-order coefficients
$B_{k_{\mathrm{1}}n_{\mathrm{1}}}^{\mathrm{I}}$, 
i.e. $B_{kr}^{\mathrm{I}}$ in eq.(\ref{be14}) can be 
calculated as in eq.(\ref{be37}). 

\section{The Calculation of the Second Order Coefficients $B_{kr}^{\mathrm{II}}$}\label{a50} 
Let us explain how to obtain the second-order coefficients, that is, 
how to solve the integral equation (\ref{be5}). 
The coefficients of first order, i.e. the family of
$B_{kr}^{\mathrm{I}}$, have been obtained as given in eq.(\ref{be37}), 
so that we can employ them to determine the second-order coefficients, i.e. the family of $B_{kr}^{\mathrm{II}}$.  

To begin with, we calculate $\Omega^{\mathrm{H}}_{kr}$ in
eq.(\ref{be25.5}) for second order; $\Omega_{kr}^{\mathrm{H}}$ for
second order corresponds to the first term on the right-hand side
of eq.(\ref{be5}). 
For second order, $\Omega^{\mathrm{H}}_{kr}$ in eq.(\ref{be25.5}) can be calculated by substituting
the family of $B_{kr}^{\mathrm{I}}$ into the expressions of $D_{k,r}$,
$E_{k,r}$ and $G_{k,r}$ in eqs.(\ref{be22}), (\ref{be23}) and (\ref{be24}); other terms do not
appear in $\Omega^{\mathrm{H}}_{kr}$ for second order. 
The results of the tedious calculation of $\Omega^{\mathrm{H}}_{kr}$ to second order finally become as follows. 
For $k=0$, $\Omega^{\mathrm{H}}_{0r}$ becomes 
\begin{eqnarray}
\Omega^{\mathrm{H}}_{0r}=0,
\label{be41r01}
\end{eqnarray}
for $r=0$ and $1$,  
\begin{eqnarray}
\Omega^{\mathrm{H}}_{02}=\frac{35}{32}\frac{1}{\sqrt{2\pi}\sigma^{2}T^{2}}\left(\frac{m}{2\kappa T}\right)^{-\frac{1}{2}}\left(\nabla T\right)^{2}
\left(b_{12}-b_{11}\right), 
\label{be41r2}
\end{eqnarray}
for $r=2$, and  
\begin{eqnarray}
\Omega^{\mathrm{H}}_{0r}=\frac{5}{16}\frac{1}{\sqrt{2\pi}\sigma^{2}T^{2}}\left(\frac{m}{2\kappa T}\right)^{-\frac{1}{2}}\left(\nabla T\right)^{2}
\left[(r^{2}+\frac{r}{2}-\frac{3}{2})b_{1r}-(2r-\frac{1}{2})b_{1,r-1}+b_{1,r-2}\right], 
\label{be41}
\end{eqnarray}
for $r \ge 3$. 
Note that values for the constants $b_{1r}$ are summarized in
Table \ref{b1r}. 
For $k=2$, $\Omega^{\mathrm{H}}_{2r}$ becomes 
\begin{eqnarray}
\Omega^{\mathrm{H}}_{20}=0,
\label{be44r0}
\end{eqnarray} 
and 
\begin{eqnarray}
\Omega^{\mathrm{H}}_{21}=-\frac{5}{8}\frac{1}{\sqrt{2\pi}\sigma^{2}T^{2}}\left(\frac{m}{2\kappa T}\right)^{-\frac{1}{2}}\left[2\left(\partial_{z} T\right)^{2}-\left(\partial_{x} T\right)^{2}-\left(\partial_{y} T\right)^{2}\right]\left(b_{12}-b_{11}\right), 
\label{be44r1}
\end{eqnarray} 
and
\begin{eqnarray}
\Omega^{\mathrm{H}}_{2r}&=&\frac{5}{16}\frac{1}{\sqrt{2\pi}\sigma^{2}T^{2}}\left(\frac{m}{2\kappa T}\right)^{-\frac{1}{2}}\left[2\left(\partial_{z} T\right)^{2}-\left(\partial_{x} T\right)^{2}-\left(\partial_{y} T\right)^{2}\right]\nonumber\\
&\times&\left[-r(r+1)b_{1,r+1}+2rb_{1r}-b_{1,r-1}\right], 
\label{be44}
\end{eqnarray} 
for $r \ge 2$. 
For $k=1$ and $k\ge 3$, we find that $\Omega^{\mathrm{H}}_{kr}$ becomes  
\begin{eqnarray}
\Omega^{\mathrm{H}}_{kr}=0,  
\label{be44.1}
\end{eqnarray} 
for any value of $r$. 

Next let us calculate $\Delta^{\mathrm{H}}_{kr}$ in eq.(\ref{be28}) for
second order. 
In order to derive $\Delta^{\mathrm{H}}_{kr}$ for second order, 
we have to calculate $W_{k,k_{\mathrm{1}},k_{\mathrm{2}}}^{n_{\mathrm{1}},
n_{\mathrm{2}},p_{\mathrm{1}},p_{\mathrm{1}}}$ and
$\Xi_{k,k_{\mathrm{1}},k_{\mathrm{2}}}^{r,n_{\mathrm{1}},n_{\mathrm{2}},p_{\mathrm{1}},p_{\mathrm{1}}}$ 
in $F_{kr}^{1}(\chi)$ of eq.(\ref{be34.5}) to second order, as was shown
in Appendix \ref{a2}.    
For even $k$, $\Delta^{\mathrm{H}}_{kr}$ to second order
results in the general form:  
\begin{eqnarray}
\Delta^{\mathrm{H}}_{kr}&=& B_{00} \sum_{n_{\mathrm{1}},0\le q\le \frac{k}{2}} B_{k-2q,n_{\mathrm{1}}}^{\mathrm{II}}  M_{k,k-2q,0}^{r,n_{\mathrm{1}},
0,0,0}+\sum_{n_{\mathrm{1}},n_{\mathrm{2}}} B_{1n_{\mathrm{1}}}^{\mathrm{I}}B_{1n_{\mathrm{2}}}^{\mathrm{I}}M_{k,1,1}^{r,n_{\mathrm{1}},
n_{\mathrm{2}},0,0}\nonumber \\
&+&\sum_{n_{\mathrm{1}},n_{\mathrm{2}}} [B^{(1)\mathrm{I}}_{1n_{\mathrm{1}}}B^{(1)\mathrm{I}}_{1n_{\mathrm{2}}}+C^{(1)\mathrm{I}}_{1n_{\mathrm{1}}}C^{(1)\mathrm{I}}_{1n_{\mathrm{2}}}]M_{k,1,1}^{r,n_{\mathrm{1}},
n_{\mathrm{2}},1,1}.   
\label{be35.8}
\end{eqnarray}
Here the set of the coefficients $B_{k-2q,n_{\mathrm{1}}}^{\mathrm{II}}B_{00}$ can be obtained from 
$W_{k,k-2q,0}^{n_{\mathrm{1}},
0,0,0}$ in eq.(\ref{be34}).  
The set of the coefficients $B_{1n_{\mathrm{1}}}^{\mathrm{I}}B_{1n_{\mathrm{2}}}^{\mathrm{I}}$ is obtained from 
$W_{k,1,1}^{n_{\mathrm{1}},
n_{\mathrm{2}},0,0}$ in eq.(\ref{be34});
$B_{1n_{\mathrm{1}}}^{\mathrm{I}}$ from $f_{\mathrm{1}}$ and $B_{1n_{\mathrm{2}}}^{\mathrm{I}}$
from $f_{\mathrm{2}}$ are the first-order coefficients obtained in eq.(\ref{be37}), so that  $B_{1n_{\mathrm{1}}}^{\mathrm{I}}B_{1n_{\mathrm{2}}}^{\mathrm{I}}$ is second order. 
Similarly, the set of the terms $B^{(1)\mathrm{I}}_{1n_{\mathrm{1}}}B^{(1)\mathrm{I}}_{1n_{\mathrm{2}}}+C^{(1)\mathrm{I}}_{1n_{\mathrm{1}}}C^{(1)\mathrm{I}}_{1n_{\mathrm{2}}}$
is obtained from $W_{k,1,1}^{n_{\mathrm{1}},
n_{\mathrm{2}},1,1}$ in eq.(\ref{be34}); 
$B^{(1)\mathrm{I}}_{1n_{\mathrm{1}}}$, $C^{(1)\mathrm{I}}_{1n_{\mathrm{1}}}$ from
$f_{\mathrm{1}}$ and 
$B^{(1)\mathrm{I}}_{1n_{\mathrm{2}}}$, $C^{(1)\mathrm{I}}_{1n_{\mathrm{2}}}$
from $f_{\mathrm{2}}$ are the first-order coefficients obtained in
\ref{first}, so that
$B^{(1)\mathrm{I}}_{1n_{\mathrm{1}}}B^{(1)\mathrm{I}}_{1n_{\mathrm{2}}}$ and
$C^{(1)\mathrm{I}}_{1n_{\mathrm{1}}}C^{(1)\mathrm{I}}_{1n_{\mathrm{2}}}$
are also second order. 
To second order, $f_{\mathrm{1}}$ of eq.(\ref{be27}) contains only $B_{00}=1$, the family
of $B_{k_{\mathrm{1}}n_{\mathrm{1}}}^{\mathrm{I}}$ obtained as eq.(\ref{be37})
and the family
of $B_{k_{\mathrm{1}}n_{\mathrm{1}}}^{\mathrm{II}}$ to be determined
here; $f_{\mathrm{2}}$ of eq.(\ref{be27}) also contains only $B_{00}=1$, the family
of $B_{k_{\mathrm{2}}n_{\mathrm{2}}}^{\mathrm{I}}$ and the family
of $B_{k_{\mathrm{2}}n_{\mathrm{2}}}^{\mathrm{II}}$ to second order. 
Thus, we can only obtain the sets of the terms in eq.(\ref{be35.8}) for
second order by using the
fact that $F_{kr}^{1}(\chi)=0$ unless
$k=|k_{\mathrm{1}}-k_{\mathrm{2}}|+2q$:   
the second and the third terms on the right-hand side of
eq.(\ref{be35.8}) do not appear for odd $k$. 
Note that it is sufficient to consider only the case for
$k_{\mathrm{1}}\ge k_{\mathrm{2}}$, as is explained in refs.\cite{burnett,kim2}, and
that  $B^{(1)\mathrm{I}}_{1n_{\mathrm{1}}}C^{(1)\mathrm{I}}_{1n_{\mathrm{2}}}$ or
$C^{(1)\mathrm{I}}_{1n_{\mathrm{1}}}B^{(1)\mathrm{I}}_{1n_{\mathrm{2}}}$ does not appear
owing to the orthogonality properties of the spherical harmonic
functions. (see Appendix \ref{a2}) 
The matrix $M_{k,k-2q,0}^{r,n_{\mathrm{1}},
0,0,0}$ in eq.(\ref{be35.8}) is thus obtained as
\begin{eqnarray}
M_{k,k-2q,0}^{r,n_{\mathrm{1}},
0,0,0}
&=&\frac{n^{2}\sigma^{2}}{2}\left(\frac{m}{2\pi\kappa T}\right)^{3}\left(\frac{m}{2\kappa T}\right)^{k-q}n_{\mathrm{1}}! (k+\frac{1}{2}) \sqrt{\pi}\nonumber\\
&\times& \int_{0}^{\pi} [\Xi_{k,k-2q,0}^{r,n_{\mathrm{1}},0,0,0}(\chi)-\Xi_{k,k-2q,0}^{r,n_{\mathrm{1}},0,0,0}(0) ]\sin\frac{\chi}{2} \cos\frac{\chi}{2} \mathrm{d}\chi, \label{be28.10}
\end{eqnarray}
using eqs.(\ref{be28}), (\ref{be34.5}) and (\ref{be34}). 
Similarly, the matrices $M_{k,1,1}^{r,n_{\mathrm{1}},
n_{\mathrm{2}},0,0}$ and $M_{k,1,1}^{r,n_{\mathrm{1}},
n_{\mathrm{2}},1,1}$ in eq.(\ref{be35.8}) are derived as 
\begin{eqnarray}
M_{k,1,1}^{r,n_{\mathrm{1}},
n_{\mathrm{2}},0,0}
&=&\frac{n^{2}\sigma^{2}}{2}\left(\frac{m}{2\pi\kappa T}\right)^{3}\left(\frac{m}{2\kappa T}\right)^{\frac{k}{2}+1}n_{\mathrm{1}}!n_{\mathrm{2}}! (k+\frac{1}{2}) \sqrt{\pi}\nonumber\\
&\times& \int_{0}^{\pi} [\Xi_{k,1,1}^{r,n_{\mathrm{1}},n_{\mathrm{2}},0,0}(\chi)-\Xi_{k,1,1}^{r,n_{\mathrm{1}},n_{\mathrm{2}},0,0}(0) ]\sin\frac{\chi}{2} \cos\frac{\chi}{2} \mathrm{d}\chi, \label{be28.11}
\end{eqnarray}
and 
\begin{eqnarray}
M_{k,1,1}^{r,n_{\mathrm{1}},
n_{\mathrm{2}},1,1}
&=&\frac{n^{2}\sigma^{2}}{2}\left(\frac{m}{2\pi\kappa T}\right)^{3}\left(\frac{m}{2\kappa T}\right)^{\frac{k}{2}+1} n_{\mathrm{1}}!n_{\mathrm{2}}! (k+\frac{1}{2}) \sqrt{\pi}\nonumber\\
&\times& \int_{0}^{\pi} [\Xi_{k,1,1}^{r,n_{\mathrm{1}},n_{\mathrm{2}},1,1}(\chi)-\Xi_{k,1,1}^{r,n_{\mathrm{1}},n_{\mathrm{2}},1,1}(0) ]\sin\frac{\chi}{2} \cos\frac{\chi}{2} \mathrm{d}\chi, 
\label{be28.12}
\end{eqnarray}
respectively. 

For even $k$, eq.(\ref{be0}) finally leads to simultaneous equations to determine
the second-order coefficients $B_{k-2q,n_{\mathrm{1}}}^{\mathrm{II}}$: 
\begin{eqnarray}
\Omega^{\mathrm{H}}_{kr}&=&
\sum_{n_{\mathrm{1}},0\le q\le \frac{k}{2}} B_{k-2q,n_{\mathrm{1}}}^{\mathrm{II}} M_{k,k-2q,0}^{r,n_{\mathrm{1}},
0,0,0}+\sum_{n_{\mathrm{1}},n_{\mathrm{2}}} B_{1n_{\mathrm{1}}}^{\mathrm{I}}B_{1n_{\mathrm{2}}}^{\mathrm{I}}M_{k,1,1}^{r,n_{\mathrm{1}},
n_{\mathrm{2}},0,0}\nonumber \\
&+&\sum_{n_{\mathrm{1}},n_{\mathrm{2}}} [B^{(1)\mathrm{I}}_{1n_{\mathrm{1}}}B^{(1)\mathrm{I}}_{1n_{\mathrm{2}}}+C^{(1)\mathrm{I}}_{1n_{\mathrm{1}}}C^{(1)\mathrm{I}}_{1n_{\mathrm{2}}}]M_{k,1,1}^{r,n_{\mathrm{1}},
n_{\mathrm{2}},1,1},   
\label{be99}
\end{eqnarray}
from eq.(\ref{be35.8}). 
The second and the third terms on the right-hand side of eq.(\ref{be99}) correspond to
$J(f^{(1)}_{\mathrm{1}},f^{(1)}_{\mathrm{2}})$ in the integral equation
(\ref{be5}). 
Thus, it had been believed that eq.(\ref{be99}) should be considered for
all even $k$ because the contribution from the right-hand side of
eq.(\ref{be5}) would not become zero even when
$\Omega^{\mathrm{H}}_{kr}$ which corresponds to the first term on the right-hand side
of eq.(\ref{be5}) is zero.\cite{burnett}
However, for $k=4$, $6$ and $8$, we have confirmed that the second and the third terms on the right
hand side of eq.(\ref{be99}) disappear, which
leads to the fact that it is not necessary to consider eq.(\ref{be99}) for $k=4$, $6$ and
$8$. 
Therefore, it is natural to expect that we do not need to consider
eq.(\ref{be99}) for even $k$ furthermore. 
Our contention that the second and the third terms on the right
hand side of eq.(\ref{be99}) will disappear for $k=2n$ with the integer
$n \ge 2$ should be demonstrated by a mathematical
proof in the future. 
It should be mentioned that the second and the third terms on the right-hand side of
eq.(\ref{be99}) do not appear for odd $k$, and $\Omega^{\mathrm{H}}_{kr}$
to second order is found to be zero for odd $k$, so that we do not need to calculate
$\Delta^{\mathrm{H}}_{kr}$ for odd $k$; it is not necessary to consider the case in
which contribution from the right-hand side of eq.(\ref{be5}) becomes
zero.\cite{resi}

Now, we need to consider eq.(\ref{be99}) only for $k=0$
and $2$. 
If $k=0$, eq.(\ref{be99}) leads to simultaneous equations to determine
the second-order coefficients $B_{0n_{\mathrm{1}}}^{\mathrm{II}}$: 
\begin{eqnarray}
\Omega^{\mathrm{H}}_{0r}&=&
\sum_{n_{\mathrm{1}}\ge 2} B_{0n_{\mathrm{1}}}^{\mathrm{II}} M_{0,0,0}^{r,n_{\mathrm{1}},
0,0,0}+\sum_{n_{\mathrm{1}},n_{\mathrm{2}}} B_{1n_{\mathrm{1}}}^{\mathrm{I}}B_{1n_{\mathrm{2}}}^{\mathrm{I}}M_{0,  1,1}^{r,n_{\mathrm{1}},
n_{\mathrm{2}},0,0}\nonumber \\
&+&\sum_{n_{\mathrm{1}},n_{\mathrm{2}}} [B^{(1)\mathrm{I}}_{1n_{\mathrm{1}}}B^{(1)\mathrm{I}}_{1n_{\mathrm{2}}}+C^{(1)\mathrm{I}}_{1n_{\mathrm{1}}}C^{(1)\mathrm{I}}_{1n_{\mathrm{2}}}]M_{0,  1,1}^{r,n_{\mathrm{1}},
n_{\mathrm{2}},1,1},   
\label{be990}
\end{eqnarray}
that is, 
\begin{eqnarray}
B_{0n_{\mathrm{1}}}^{\mathrm{II}}=\sum_{r\ge 2} \left\{\Omega^{\mathrm{H}}_{0r}-\sum_{n_{\mathrm{1}},n_{\mathrm{2}}} B_{1n_{\mathrm{1}}}^{\mathrm{I}}B_{1n_{\mathrm{2}}}^{\mathrm{I}}M_{0,  1,1}^{r,n_{\mathrm{1}},
n_{\mathrm{2}},0,0} \right.\nonumber \\
\left. -\sum_{n_{\mathrm{1}},n_{\mathrm{2}}} [B^{(1)\mathrm{I}}_{1n_{\mathrm{1}}}B^{(1)\mathrm{I}}_{1n_{\mathrm{2}}}+C^{(1)\mathrm{I}}_{1n_{\mathrm{1}}}C^{(1)\mathrm{I}}_{1n_{\mathrm{2}}}]M_{0,  1,1}^{r,n_{\mathrm{1}},
n_{\mathrm{2}},1,1}\right\}(M_{0,0,0}^{r,n_{\mathrm{1}},
0,0,0})^{-1}, \label{be131}
\end{eqnarray}
with $\Omega^{\mathrm{H}}_{0r}$ in eqs.(\ref{be41r01}), (\ref{be41r2}) and (\ref{be41}). 
We should derive the second-order coefficients
$B_{0n_{\mathrm{1}}}^{\mathrm{II}}$ only for $n_{\mathrm{1}}\ge 2$, because $B_{00}=1$ and $B_{01}=0$ from
eq.(\ref{be18}). 
We have calculated the matrix $M_{0,0,0}^{r,n_{\mathrm{1}},
0,0,0}$ for $2 \le r \le 6$ and $2 \le n_{\mathrm{1}} \le 6$  from
eq.(\ref{be28.10}), and we have confirmed that 
$M_{0,0,0}^{r,n_{\mathrm{1}},
0,0,0}$ vanishes for $r=0, 1$ and $2 \le n_{\mathrm{1}} \le 6$. 
We have also calculated the matrices $M_{0,
1,1}^{r,n_{\mathrm{1}},n_{\mathrm{2}},0,0}$ for $2\le r \le 6$, $1\le
n_{\mathrm{1}} \le 7$ and $1\le n_{\mathrm{2}} \le 7$ from
eq.(\ref{be28.11}), and we have confirmed
$M_{0,
1,1}^{r,n_{\mathrm{1}},n_{\mathrm{2}},0,0}$ vanishes for $r=0,
1$, $1\le n_{\mathrm{1}} \le 7$ and $1\le n_{\mathrm{2}} \le 7$. 
Our results for $M_{0,0,0}^{r,n_{\mathrm{1}},
0,0,0}$ for $2 \le r \le 6$ and $2 \le n_{\mathrm{1}} \le 6$  and $M_{0,
1,1}^{r,n_{\mathrm{1}},n_{\mathrm{2}},0,0}$ for $2\le r \le 6$, $1\le n_{\mathrm{1}}
\le 7$ and $1\le n_{\mathrm{2}} \le 7$ are given in ref.\cite{address}. 
The matrices $M_{0,
1,1}^{r,n_{\mathrm{1}},n_{\mathrm{2}},1,1}$ for $2\le r
\le 6$, $1\le n_{\mathrm{1}} \le 7$ and $1\le n_{\mathrm{2}} \le 7$, 
which can be calculated from eq.(\ref{be28.12}), are also obtained from
the symmetric relation $M_{0, 1,1}^{r,n_{\mathrm{1}},
n_{\mathrm{2}},0,0}=M_{0,  1,1}^{r,n_{\mathrm{1}},
n_{\mathrm{2}},1,1}$ arising from properties of the
spherical harmonic function. 
Finally, we can determine the second-order coefficients
$B_{0n_{\mathrm{1}}}^{\mathrm{II}}$ in $f_{\mathrm{1}}$,
i.e. $B_{0r}^{\mathrm{II}}$ in eq.(\ref{be13}) as in eq.(\ref{be42}). 

If $k=2$, eq.(\ref{be99}) leads to simultaneous equations to determine
the second-order coefficients $B_{2n_{\mathrm{1}}}^{\mathrm{II}}$: 
\begin{eqnarray}
\Omega^{\mathrm{H}}_{2r}&=&
\sum_{n_{\mathrm{1}}\ge 0} [B_{2n_{\mathrm{1}}}^{\mathrm{II}} M_{2,2,0}^{r,n_{\mathrm{1}},
0,0,0}+B_{0n_{\mathrm{1}}}^{\mathrm{II}} M_{2,0,0}^{r,n_{\mathrm{1}},
0,0,0}]+\sum_{n_{\mathrm{1}},n_{\mathrm{2}}} B_{1n_{\mathrm{1}}}^{\mathrm{I}}B_{1n_{\mathrm{2}}}^{\mathrm{I}}M_{2,  1,1}^{r,n_{\mathrm{1}},
n_{\mathrm{2}},0,0}\nonumber \\
&+&\sum_{n_{\mathrm{1}},n_{\mathrm{2}}} [B^{(1)\mathrm{I}}_{1n_{\mathrm{1}}}B^{(1)\mathrm{I}}_{1n_{\mathrm{2}}}+C^{(1)\mathrm{I}}_{1n_{\mathrm{1}}}C^{(1)\mathrm{I}}_{1n_{\mathrm{2}}}]M_{2,  1,1}^{r,n_{\mathrm{1}},
n_{\mathrm{2}},1,1},   
\label{be992}
\end{eqnarray}
that is, 
\begin{eqnarray}
B_{2n_{\mathrm{1}}}^{\mathrm{II}}=\sum_{r\ge 0} \left\{\Omega^{\mathrm{H}}_{2r}-\sum_{n_{\mathrm{1}},n_{\mathrm{2}}} B_{1n_{\mathrm{1}}}^{\mathrm{I}}B_{1n_{\mathrm{2}}}^{\mathrm{I}}M_{2,  1,1}^{r,n_{\mathrm{1}},
n_{\mathrm{2}},0,0}\right. \nonumber \\
\left. -\sum_{n_{\mathrm{1}},n_{\mathrm{2}}} [B^{(1)\mathrm{I}}_{1n_{\mathrm{1}}}B^{(1)\mathrm{I}}_{1n_{\mathrm{2}}}+C^{(1)\mathrm{I}}_{1n_{\mathrm{1}}}C^{(1)\mathrm{I}}_{1n_{\mathrm{2}}}]M_{2,  1,1}^{r,n_{\mathrm{1}},
n_{\mathrm{2}},1,1}\right\}(M_{2,2,0}^{r,n_{\mathrm{1}},
0,0,0})^{-1}, \label{be123}
\end{eqnarray}
with $\Omega^{\mathrm{H}}_{2r}$ in eqs.(\ref{be44r0}), (\ref{be44r1}),
and (\ref{be44}). 
Note that we have confirmed $M_{2,0,0}^{r,n_{\mathrm{1}},
0,0,0}$ in eq.(\ref{be992}) becomes zero, which
had been also confirmed by Burnett\cite{burnett}. 
We should derive the second-order coefficients
$B_{2n_{\mathrm{1}}}^{\mathrm{II}}$ for $n_{\mathrm{1}}\ge 0$. 
We have calculated the matrix $M_{2,2,0}^{r,n_{\mathrm{1}},
0,0,0}$ for $0\le r \le 6$ and $0 \le n_{\mathrm{1}} \le 6$  from
eq.(\ref{be28.10}), and also the matrices $M_{2,
1,1}^{r,n_{\mathrm{1}},n_{\mathrm{2}},0,0}$ for $0\le r
\le 6$ , $1\le n_{\mathrm{1}}\le 7$ and $1\le n_{\mathrm{2}}\le 7$ from eq.(\ref{be28.11}). 
Our results are given in ref.\cite{address}. 
The matrices $M_{2, 1,1}^{r,n_{\mathrm{1}},n_{\mathrm{2}},1,1}$ for $0\le r
\le 6$, $1\le n_{\mathrm{1}} \le 7$ and $1 \le n_{\mathrm{2}} \le 7$, which can be calculated from eq.(\ref{be28.12}), are also obtained from
the symmetric relation $M_{2, 1,1}^{r,n_{\mathrm{1}},
n_{\mathrm{2}},0,0}=-2M_{2,  1,1}^{r,n_{\mathrm{1}},
n_{\mathrm{2}},1,1}$ arising from properties of the
spherical harmonic function.\cite{burnett}   
The second-order coefficients
$B_{k_{\mathrm{1}}n_{\mathrm{1}}}^{\mathrm{II}}$ in $f_{\mathrm{1}}$,
i.e. $B_{kr}^{\mathrm{II}}$ in eq.(\ref{be14}) can be written
 in the final form shown in eq.(\ref{be45}). 

\section{Test of the Nonequilibrium Steady-State Thermodynamics}\label{a6}
\begin{figure}[htbp]
\begin{center}
\includegraphics[width=8.cm]{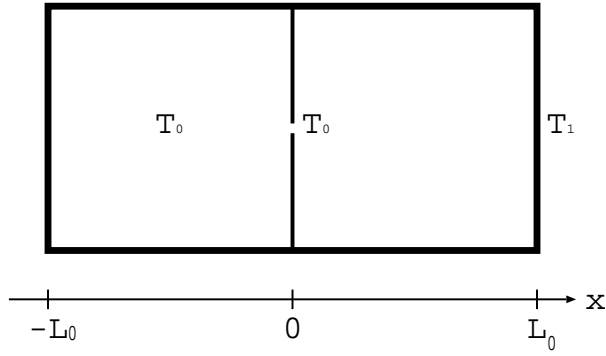} 
\caption{The cell on the left-hand side is at equilibrium at temperature $T_{0}$.
The cell on the right-hand side is in a
 nonequilibrium state under a temperature gradient along the $x$-axis caused by the right
 wall (at $x=L_{0}$)
 at temperature $T_{1}$ and the thin mid-wall (at $x=0$) at temperature
 $T_{0}$, of thickness less than or
 equal to the mean free path $l$ of the dilute gases. 
Both cells are filled with dilute gases and connected by a small
 hole of diameter $d$ on the thin mid-wall.  
The diameter of the small hole $d$ is much smaller than $l$, i.e. $d \ll l$.  
Molecules which have passed through the small hole are relaxed into the
state of the cell they go into after a few interactions, so that they do
not affect the macroscopic state of that cell. 
The mean mass flux at the small hole is zero in the nonequilibrium
 steady state. }
\label{sst}
\end{center}
\end{figure}

We examine SST from the microscopic viewpoint, that
is, by applying the velocity distribution function for
the steady-state Boltzmann equation for both hard-core molecules and
Maxwell molecules to the cell on the right-hand side in a simple nonequilibrium steady-state system
illustrated in Fig.\ref{sst}. 
We also use the velocity distribution function for the steady-state BGK
equation to second order in the test of SST. 
This simple nonequilibrium steady-state system is inspired by SST suggested by Sasa and Tasaki\cite{sst}. 
They predicted the following results. 
When the system shown in Fig.\ref{sst} is in a steady-state, the
osmosis, defined as the difference between the pressure $P_{xx}$ of the cell in the nonequilibrium steady
state and the value $P_{0}$ of the
cell at equilibrium, namely 
\begin{eqnarray} 
\Delta P&\equiv&P_{xx}-P_{0}, \label{be151} 
\end{eqnarray}
always becomes positive. 
Additionally, there is a relation 
\begin{eqnarray}  
\frac{n(0)}{n_{0}}=\left(\frac{\partial P_{xx}}{\partial P_{0}}\right)_{T_{0},J_{x}}, \label{be53}
\end{eqnarray}
connecting $n(0)$, $P_{xx}$, $n_{0}$ and $P_{0}$, 
where $n(0)$ is the density of the cell in the nonequilibrium steady
state around the hole and $n_{0}$ is that of the
cell at equilibrium. 

We consider the nonequilibrium steady-state where
the mean mass flux at the hole is zero:
\begin{eqnarray}  
\int^{\infty}_{0}\mathrm{d}c_{x}\int^{\infty}_{-\infty}\mathrm{d}^{2}c_{\bot}mc_{x}f_{0}+\int^{0}_{-\infty}\mathrm{d}c_{x}\int^{\infty}_{-\infty}\mathrm{d}^{2}c_{\bot}mc_{x}f|_{x=0}=0, \label{be48}
\end{eqnarray}
where $c_{\bot}$ represents the components of the velocity which are
orthogonal to $c_{x}$, i.e. $c_{y}$ and $c_{z}$. 
Note that we consider the mean mass flux at the hole, so that we put $x=0$
in $f$. 
$f_{0}=n_{0}(m/2\pi \kappa T_{0})^{3/2}e^{-m{\bf v}^{2}/2\kappa T_{0}}$ is the velocity distribution function of the cell on the
left-hand side at equilibrium at
temperature $T_{0}$ and density $n_{0}$. 
From eq.(\ref{be48}), the relation between $n(0)$ and $n_{0}$ is obtained as 
\begin{eqnarray}  
n(0)=n_{0}[1+\lambda_{n}\frac{mJ_{x}^{2}}{n_{0}^{2}\kappa^{3}T_{0}^{3}}], \label{be50}
\end{eqnarray}
to second order.  
The value for the constant $\lambda_{n}$ is given in Table \ref{macro2}. 
The density of the cell in the nonequilibrium steady
state around the hole $n(0)$ is greater than that of the
cell at equilibrium $n_{0}$ regardless of the sign of $J_{x}$. 
We emphasize that $\lambda_{n}$ could not be calculated if we did not derive the explicit form of the velocity
distribution function $f$ to second order. 
We have adopted the boundary condition around the hole, $T(0)=T_{0}$, in 
order to examine SST.  
Although it is believed that the Knudsen layer effect, i.e. the slip effect is dominant
around the 'wall'\cite{cercignani1,maxwell,cercignani}, for reasons
of simplification, we do not consider the slip effect around the 'hole' in this paper. 
If the slip effect is not dominant, the density of the cell in the nonequilibrium steady
state will always be larger than that of the
cell at equilibrium, regardless of the sign of $J_{x}$. 
This can be tested by experiments on the
nonequilibrium steady-state system shown in Fig.\ref{sst}. 

We also calculate the osmosis $\Delta P$ in eq.(\ref{be151}) as 
\begin{eqnarray} 
\Delta P&=&\lambda_{\Delta P}\frac{mJ_{x}^{2}}{n_{0}\kappa^{2}T_{0}^{2}}, \label{be52}
\end{eqnarray}
to second order, using eqs.(\ref{be51}), (\ref{be50}) and $P_{0}=n_{0}\kappa T_{0}$. 
Values for $\lambda_{\Delta P}$ are given in Table \ref{macro2}. 
We have found that $\lambda_{\Delta P}$ is always positive, which agrees with the
prediction by SST\cite{sst}. 
Furthermore, we are able to test the relation (\ref{be53}). 
Though substitution of eqs.(\ref{be51}) and (\ref{be50}) into eq.(\ref{be53})
leads to the relation $\lambda_{P}^{xx}/\lambda_{n}=-2$, our numerical results conflict with this
relation: our results give $\lambda_{P}^{xx}/\lambda_{n}=-0.5604$
for $7$th approximation $b_{0r}$ and
$b_{2r}$ for hard-core molecules and $\lambda_{P}^{xx}/\lambda_{n}=0$
for both Maxwell molecules and the steady-state BGK equation. 
We have also confirmed that the relation
$\lambda_{P}^{xx}/\lambda_{n}=-2$ predicted in SST\cite{sst} is not
modified if the boundary condition around the hole can be written in the form: 
\begin{eqnarray}  
T(0)=T_{0}[1+\lambda_{T}\frac{mJ_{x}^{2}}{n_{0}^{2}\kappa^{3}T_{0}^{3}}], \label{be57}
\end{eqnarray}
where the constant $\lambda_{T}$ represents the difference
between the temperature of molecules around the hole and that of the
mid-wall. 
Our boundary condition $T(0)=T_{0}$ corresponds to putting
$\lambda_{T}=0$ for any kinetic models. 
It can be concluded that, although we regard a state in which the mean mass
flux at the hole is zero to be a nonequilibrium steady state, as in
eq.(\ref{be48}), this state has yet to be interpreted
phenomenologically.  

On the other hand, by virtue of the derivation of the explicit form of the velocity
distribution function $f$ to second order, we can also calculate the $x$ component of the heat
flux at the hole as 
\begin{eqnarray}  
J^{*}&\equiv&\left(\frac{2\kappa T}{m}\right)^{3}\int^{\infty}_{0}\mathrm{d}c_{x}\int^{\infty}_{-\infty}\mathrm{d}^{2}c_{\bot}\frac{m{\bf c}^{2}}{2}c_{x}f_{0}+\left(\frac{2\kappa T}{m}\right)^{3}\int^{0}_{-\infty}\mathrm{d}c_{x}\int^{\infty}_{-\infty}\mathrm{d}^{2}c_{\bot}\frac{m{\bf c}^{2}}{2}c_{x}f|_{x=0} \nonumber\\
&=&\frac{J_{x}}{2}+\lambda_{J^{*}}\frac{mJ_{x}^{2}}{n_{0}\kappa^{2}T_{0}^{2}}\left(\frac{2\kappa T_{0}}{\pi m}\right)^{\frac{1}{2}},\label{be55}
\end{eqnarray} 
to second order using eq.(\ref{be50}).
Numerical values for the constant $\lambda_{J^{*}}$ are given in Table \ref{macro2}. 
The first-order term on the right-hand side of eq.(\ref{be55}) is from Fourier's
law, while the second-order term also appears on the right-hand side of
eq.(\ref{be55}), though the second-order heat flux ${\bf J}^{(2)}$ does not
exist in the cell in the nonequilibrium steady state. 
This fact indicates that $T(0)$ is not appropriate
for the \textit{nonequilibrium temperature}, if $J^{*}$ suggests the
existence of the \textit{nonequilibrium temperature}.

\newpage

\begin{table}[htbp]
\begin{center}
\caption{\label{b1r}Numerical constants $b_{1r}$ in
 eq.(\ref{be37})}
\begin{tabular}{@{\hspace{\tabcolsep}\extracolsep{\fill}}cccccc} \hline 
{$r$}&{$r \le 4$}&{$r \le 5$}&{$r \le 6$}&{$r \le 7$}&{Maxwell's $b_{1r}$} \\ \hline
{$1$}&{$1.025$}&{$1.025$}&{$1.025$}&{$1.025$}&{$1$} \\ \hline
{$2$}&{$4.881\times 10^{-2}$}&{$4.889\times 10^{-2}$}&{$4.891\times 10^{-2}$}&{$4.892\times 10^{-2}$}&{$0$} \\ \hline
{$3$}&{$3.639\times 10^{-3}$}&{$3.698\times 10^{-3}$}&{$3.711\times 10^{-3}$}&{$3.715\times 10^{-3}$}&{$0$}
 \\ \hline
{$4$}&{$2.526\times 10^{-4}$}&{$2.838\times 10^{-4}$}&{$2.905\times 10^{-4}$}&{$2.922\times 10^{-4}$}&{$0$} \\ \hline
{$5$}&{$-$}&{$1.855\times 10^{-5}$}&{$2.123\times 10^{-5}$}&{$2.187\times 10^{-5}$}&{$0$} \\ \hline
{$6$}&{$-$}&{$-$}&{$1.284\times 10^{-6}$}&{$1.492\times 10^{-6}$}&{$0$} \\ \hline
{$7$}&{$-$}&{$-$}&{$-$}&{$8.322\times 10^{-8}$}&{$0$} \\ \hline 
\end{tabular}
\end{center}
\end{table}

\begin{table}[htbp]
\begin{center}
\caption{\label{b0r}Numerical constants $b_{0r}$ in
 eq.(\ref{be42})}
\begin{tabular}{@{\hspace{\tabcolsep}\extracolsep{\fill}}cccccc} \hline 
{$r$}&{$r \le 4$}&{$r \le 5$}&{$r \le 6$}&{$r \le 7$}&{Maxwell's $b_{0r}$} \\ \hline
{$2$}&{$4.434\times 10^{-1}$}&{$4.390\times 10^{-1}$}&{$4.381\times 10^{-1}$}&{$4.380\times 10^{-1}$}&{$\frac{825}{1024}$}
 \\ \hline
{$3$}&{$-4.935\times 10^{-2}$}&{$-5.342\times 10^{-2}$}&{$-5.413\times 10^{-2}$}&{$-5.429\times 10^{-2}$}&{$-\frac{25}{256}$} \\ \hline
{$4$}&{$-$}&{$-3.581\times 10^{-3}$}&{$-4.007\times 10^{-3}$}&{$-4.098\times 10^{-3}$}&{$0$} \\ \hline
{$5$}&{$-$}&{$-$}&{$-2.779\times 10^{-4}$}&{$-3.184\times 10^{-4}$}&{$0$} \\ \hline
{$6$}&{$-$}&{$-$}&{$-$}&{$-2.087\times 10^{-5}$}&{$0$} \\ \hline 
\end{tabular}
\end{center}
\end{table}

\begin{table}[htbp]
\begin{center}
\caption{\label{b2r}Numerical constants $b_{2r}$ in
 eq.(\ref{be45})}
\begin{tabular}{@{\hspace{\tabcolsep}\extracolsep{\fill}}cccccc} \hline 
{$r$}&{$r \le 4$}&{$r \le 5$}&{$r \le 6$}&{$r \le 7$}&{Maxwell's $b_{2r}$}  \\ \hline
{$0$}&{$-3.353\times 10^{-2}$}&{$-3.327\times 10^{-2}$}&{$-3.322\times 10^{-2}$}&{$-3.320\times 10^{-2}$}&{$0$} \\ \hline
{$1$}&{$-1.285\times 10^{-1}$}&{$-1.278\times 10^{-1}$}&{$-1.277\times 10^{-1}$}&{$-1.276\times 10^{-1}$}&{$\frac{75}{896}$} \\ \hline
{$2$}&{$6.320\times 10^{-1}$}&{$6.394\times 10^{-2}$}&{$6.410\times 10^{-2}$}&{$6.414\times 10^{-2}$}&{$\frac{125}{1536}$}
 \\ \hline
{$3$}&{$4.884\times 10^{-3}$}&{$5.395\times 10^{-3}$}&{$5.496\times 10^{-3}$}&{$5.521\times 10^{-3}$}&{$0$} \\ \hline
{$4$}&{$-$}&{$3.609\times 10^{-4}$}&{$4.101\times 10^{-4}$}&{$4.214\times 10^{-4}$}&{$0$} \\ \hline
{$5$}&{$-$}&{$-$}&{$2.685\times 10^{-5}$}&{$3.106\times 10^{-5}$}&{$0$} \\ \hline
{$6$}&{$-$}&{$-$}&{$-$}&{$1.861\times 10^{-6}$}&{$0$} \\ \hline 
\end{tabular}
\end{center}
\end{table}

\begin{table}[htbp]
\begin{center}
\caption{\label{macro1}Numerical constants for the macroscopic
 quantities I: the $i$th approximation quantities for hard-core molecules and the exact values for
 Maxwell molecules and the steady-state BGK equation. }
\begin{tabular}{@{\hspace{\tabcolsep}\extracolsep{\fill}}cccccc} \hline 
{$i$th}&{$\lambda_{P}^{xx}$}&{$\lambda_{P}^{yy}$}&{$\lambda_{T_{x}}$}&{$\lambda_{T_{y}}$}&{$\lambda_{S}$}\\ \hline
{$4$th}&{$-4.647\times 10^{-2}$}&{$2.324\times 10^{-2}$}&{$-2.324\times 10^{-2}$}&{$1.162\times 10^{-2}$}&{$-2.034\times 10^{-1}$} \\ \hline
{$5$th}&{$-4.610\times 10^{-2}$}&{$2.305\times 10^{-2}$}&{$-2.305\times 10^{-2}$}&{$1.153\times 10^{-2}$}&{$-2.035\times 10^{-1}$} \\ \hline
{$6$th}&{$-4.602\times 10^{-2}$}&{$2.301\times 10^{-2}$}&{$-2.301\times 10^{-2}$}&{$1.151\times 10^{-2}$}&{$-2.035\times 10^{-1}$} \\ \hline
{$7$th}&{$-4.600\times 10^{-2}$}&{$2.300\times 10^{-2}$}&{$-2.300\times 10^{-2}$}&{$1.150\times 10^{-2}$}&{$-2.035\times 10^{-1}$} \\ \hline
{Maxwell}&{$0$}&{$0$}&{$0$}&{$0$}&{$-\frac{1}{5}$}\\ \hline
{BGK equation}&{$0$}&{$0$}&{$0$}&{$0$}&{$-\frac{1}{5}$}\\ \hline 
\end{tabular}
\end{center}
\end{table}

\begin{table}[htbp]
\begin{center}
\caption{\label{macro2}Numerical constants for the macroscopic
 quantities II: the $i$th approximation quantities for hard-core molecules and the exact values for
 Maxwell molecules and the steady-state BGK equation. 
The values of $\lambda_{n}$, $\lambda_{\Delta P}$ and $\lambda_{J^{*}}$
for hard-core molecules has not yet converged to
$4$th approximation $b_{0r}$ and $b_{2r}$ values. 
The ratios of the $7$th to the $6$th approximation
$\lambda_{n}$, $\lambda_{\Delta P}$ and $\lambda_{J^{*}}$ are
$1.011$, $1.023$ and $1.004$, respectively, so that the errors included
 in the $7$th
approximation $\lambda_{n}$, $\lambda_{\Delta P}$ and $\lambda_{J^{*}}$
appear to be less than about one or two percent. 
Note that $n(0)=n_{0}$ to first order, and that osmosis does not appear
to first order. }
\begin{tabular}{@{\hspace{\tabcolsep}\extracolsep{\fill}}cccc} \hline 
{$i$th}&{$\lambda_{n}$}&{$\lambda_{\Delta P}$}&{$\lambda_{J^{*}}$} \\ \hline
{$4$th}&{$9.255\times 10^{-2}$}&{$4.608\times 10^{-2}$}&{$-3.237\times 10^{-1}$} \\ \hline
{$5$th}&{$8.528\times 10^{-2}$}&{$3.917\times 10^{-2}$}&{$-3.109\times 10^{-1}$} \\ \hline
{$6$th}&{$8.296\times 10^{-2}$}&{$3.694\times 10^{-2}$}&{$-3.073\times 10^{-1}$} \\ \hline
{$7$th}&{$8.210\times 10^{-2}$}&{$3.609\times 10^{-2}$}&{$-3.060\times 10^{-1}$} \\ \hline
{Maxwell}&{$\frac{71}{1575}$}&{$\frac{71}{1575}$}&{$-\frac{41}{105}$}\\ \hline
{BGK equation}&{$\frac{2}{25}$}&{$\frac{2}{25}$}&{$-\frac{11}{25}$}\\ \hline 
\end{tabular}
\end{center}
\end{table}

\newpage

\end{document}